\documentclass[preprintnumbers, floatfix,letterpaper,aps,prd,epsfig,nofootinbib,
longbibliography,
twocolumn
]{revtex4-1}
\usepackage{bm,graphicx,dcolumn,epstopdf,epsf, latexsym,mathbbol, amssymb,amsmath, color, slashed, mathrsfs,mathcomp,simplewick}
\usepackage[center]{subfigure}
\usepackage{multirow}
\usepackage{makecell}
\usepackage{ytableau}
\usepackage{booktabs}
\usepackage[colorlinks,linkcolor=blue,citecolor=blue,urlcolor=blue]{hyperref}
\begin{document}
\allowdisplaybreaks
%%%%%%%%%%%%%%%%%%%%%%%%
 \newcommand{\bq}{\begin{equation}}
 \newcommand{\eq}{\end{equation}}
 \newcommand{\bqn}{\begin{eqnarray}}
 \newcommand{\eqn}{\end{eqnarray}}
 \newcommand{\nb}{\nonumber}
 \newcommand{\lb}{\label}
 \newcommand{\f}{\frac}
 \newcommand{\p}{\partial}
%%%%%%%%%%%%%%%%%%%%%%%%%
\newcommand{\PRL}{Phys. Rev. Lett.}
\newcommand{\PLB}{Phys. Lett. B}
\newcommand{\PRD}{Phys. Rev. D}
\newcommand{\CQG}{Class. Quantum Grav.}
\newcommand{\JCAP}{J. Cosmol. Astropart. Phys.}
\newcommand{\JHEP}{J. High. Energy. Phys.}
\newcommand{\bea}{\begin{eqnarray}}
\newcommand{\ena}{\end{eqnarray}}
\newcommand{\beqa}{\begin{eqnarray}}
\newcommand{\eeqa}{\end{eqnarray}}
\newcommand{\red}{\textcolor{red}}

\newlength\scratchlength
\newcommand\s[2]{% #1 - (up)scale-factor; #2 = content
  \settoheight\scratchlength{\mathstrut}%
  \scratchlength=\number\numexpr\number#1-1\relax\scratchlength
%  \scratchlength=\number\numexpr\number#1-1\relax\ht\strutbox
  \lower.5\scratchlength\hbox{\scalebox{1}[#1]{$#2$}}%
}

 %%%%%%%%%%%%%%%%%%%%%%%%

\title{Gravitational Wave Signatures from Periodic Orbits around a non-commutative inspired black hole surrounded by quintessence}

\author{Fazlay Ahmed${}^{a, b}$}
\email{fazleyahmad12@gmail.com}

\author{Qiang Wu${}^{a, b}$}
\email{wuq@zjut.edu.cn}

\author{Sushant~G.~Ghosh${}^{c, d}$}
\email{sghosh2@jmi.ac.in}

\author{Tao Zhu${}^{a, b}$}
\email{Corresponding author: zhut05@zjut.edu.cn}

\affiliation{${}^{a}$Institute for Theoretical Physics \& Cosmology, Zhejiang University of Technology, Hangzhou, 310023, China\\
${}^{b}$ United Center for Gravitational Wave Physics (UCGWP),  Zhejiang University of Technology, Hangzhou, 310023, China\\
${}^{c}$Centre for Theoretical Physics,
Jamia Millia Islamia, New Delhi 110025, India\\
${}^{d}$ Astrophysics and Cosmology Research Unit,
School of Mathematics, Statistics and Computer Science,
University of KwaZulu-Natal, Private Bag 54001, Durban 4000, South Africa}

\date{\today}

\begin{abstract}

We study gravitational wave emission from periodic orbits of a test particle around a noncommutative-inspired black hole surrounded by quintessence. Using the zoom-whirl taxonomy, which is characterized by three topological numbers $(z, w, v)$, we classify these orbits and calculate several representative gravitational waveforms for certain periodic orbits. We find that the noncommutative parameter $\Theta$ and the quintessence field significantly modify both the orbital structure and the emitted waveforms. In particular, increasing $\Theta$ leads to a phase shift and a change in amplitude in the waveform, while higher zoom numbers produce more complicated substructures. The characteristic strain spectra peak in the millihertz range, lying within the sensitivity band of the LISA detector. Moreover, the presence of the quintessence field introduces significant modifications to these waveforms, imprinting measurable deviations that could be tested or constrained by future space-based gravitational wave detectors.  These results suggest that future space-based gravitational wave missions could probe or constrain noncommutative effects in strong gravitational fields.

\end{abstract}

%\pacs{98.80.Cq, 98.80.Qc, 04.50.Kd, 04.60.Bc}

\maketitle

\section{Introduction}
\renewcommand{\theequation}{1.\arabic{equation}} \setcounter{equation}{0}

The advent of gravitational wave astronomy, marked by the groundbreaking detection of gravitational waves by LIGO and Virgo in 2015, has opened a new frontier in our exploration of the universe \cite{LIGOScientific:2016aoc, LIGOScientific:2016vbw, LIGOScientific:2016vlm, LIGOScientific:2016emj}. These spacetime ripples, predicted by Einstein's general theory of relativity, offer a unique observational window into the most energetic and violent cosmic events, such as binary black holes and binary neutron star mergers. Beyond these cataclysmic phenomena, the study of particle trajectories around black holes provides a powerful theoretical framework for probing the intricate dynamics of strong gravitational fields. Among these trajectories, periodic orbits are particularly significant because of their role in addressing fundamental challenges in astrodynamics. The analysis of periodic orbits sheds light not only on the stability of celestial systems and the complex interactions between black holes and their surrounding matter, but also provides fundamental insights into generic orbital dynamics \cite{Levin:2008mq, Levin:2009sk, Misra:2010pu, Babar:2017gsg}. All generic orbits around black holes can be considered as minor deviations from periodic orbits \cite{Levin:2008mq}. The study of periodic orbits and their gravitational wave emissions is also of particular interest because of their potential observational applications in future space-based gravitational wave detectors. 

Black holes with stellar mass or neutron stars are often found in close orbits around supermassive black holes (SMBHs). Such binary systems are known as the extreme mass ratio inspiral (EMRI), being one of the most critical targets of future space-based gravitational detectors, such as Taiji \cite{Hu:2017mde}, Tianqin \cite{TianQin:2015yph, Gong:2021gvw}, LISA \cite{Danzmann:1997hm, Schutz:1999xj, Gair:2004iv, LISA:2017pwj, Maselli:2021men}, etc. Analyzing the signals of the gravitational waveforms allows for precise measurements of the compact object’s orbital motion and the black hole’s gravitational field, offering key insights into the evolution of the Universe and strong-field gravity \cite{Bian:2025ifp, Ni:2024acg}. Given that the energy carried away by the orbital motion of the lower mass object is an exceedingly small fraction of the total energy of the system, the time it takes for the smaller mass object to spiral around the supermassive black hole can span several years. During this process, the orbital dynamics of the smaller-mass object can be well approximated by periodic orbits.  

A systematic classification of periodic orbits for massive particles provides valuable insight into the dynamical processes involved in black hole mergers \cite{Levin:2008mq}. The primary concept of this classification scheme is that a dynamic system can be understood by studying its periodic orbits. To be exact, there are three topological integers indexing all closed orbits around a black hole, representing scaling ($z$), rotation ($\omega$), and vertex ($\nu$) behaviors, respectively. Under this taxonomy, extensive research has been carried out on periodic orbits within various black hole spacetimes, to mention a few, including those of Schwarzschild and Kerr \cite{Levin:2008ci, Levin:2009sk, Bambhaniya:2020zno, Rana:2019bsn}, charged black hole \cite{Misra:2010pu}, naked singularities \cite{Babar:2017gsg}, Kerr-Sen black holes \cite{Liu:2018vea}, and hairy black holes in Horndeski's theory \cite{Lin:2023rmo}. For the studies of periodic orbits in other black holes, see refs.~\cite{Yao:2023ziq, Lin:2022llz, Chan:2025ocy, Wang:2022tfo, Lin:2023eyd, Haroon:2025rzx, Habibina:2022ztd, Zhang:2022psr, Lin:2022wda, Gao:2021arw, Lin:2021noq, Deng:2020yfm, Tu:2023xab, Zhou:2020zys, Gao:2020wjz, Deng:2020hxw, Azreg-Ainou:2020bfl, Wei:2019zdf, Pugliese:2013xfa,Zhang:2022zox, Healy:2009zm, Wang:2025wob, Alloqulov:2025bxh, Wei:2025qlh} and references therein. The gravitational wave emissions from the periodic orbits of a large number of black hole spacetimes have also been studied; see refs. \cite{Tu:2023xab, Yang:2024lmj, Shabbir:2025kqh, Junior:2024tmi, Zhao:2024exh, Jiang:2024cpe, Yang:2024cnd, Meng:2024cnq, Li:2024tld, QiQi:2024dwc, Haroon:2025rzx, Alloqulov:2025ucf, Wang:2025hla, Lu:2025cxx, Zare:2025aek, Gong:2025mne, Li:2025sfe, Choudhury:2025qsh, Chen:2025aqh, Deng:2025wzz, Li:2025eln, Zahra:2025tdo} and references therein. 

In this paper, we investigate the gravitational wave emission from the periodic orbital motion of a test particle around a black hole surrounded by quintessence in the context of noncommutative theory. The primary purpose of this article is to study the periodic orbital behaviors of a particle surrounding a black hole in non-commutative geometry and their corresponding gravitational wave radiations. We explore how noncommutative effects affect the behavior of orbits and calculate the corresponding gravitational wave radiation. 
The article is constructed as follows. In Section~\ref{section2}, we present a brief review of non-commutative black hole solutions. Then, in Section~\ref{section3}, we discuss the geodesics of a massive test particle around black holes in noncommutative geometry and study the corresponding periodic orbits. In Section~\ref{section4}, we calculate the gravitational wave radiation of periodic orbits around black holes. Conclusion and discussion are presented in Section~\ref {section5}.

\section{Black holes in non-commutative inspired geometry}\label{section2}
\renewcommand{\theequation}{2.\arabic{equation}} \setcounter{equation}{0}

In this section, we give a concise review of static, spherically symmetric black hole solutions in non-commutative–inspired geometry in the presence of a surrounding quintessence field. We begin with the line element
\begin{align}\label{metric1}
ds^2=-f(r)\,dt^2+\frac{1}{f(r)}\,dr^2+r^2 d\Omega^2,
\end{align}
with
\begin{equation}\label{fr}
    f(r)=1-\frac{2M}{r}-\frac{p}{r^{3\omega+1}}\,,
\end{equation}
where $M$ denotes the black hole mass, $\omega$ is the quintessence equation-of-state parameter (commonly taken in the range $-1<\omega<-1/3$, and $p$ is a positive normalization constant characterising the quintessence distribution. The last term in the above Eq.~\eqref{fr} follows the Kiselev ansatz for a static, anisotropic fluid that models quintessence; in this construction, the energy density of the quintessence fluid scales with radius and is related to the parameter $p$ (see e.g. \citep{Kiselev:2002dx}). In the limit $p\to0$ the metric (\ref{metric1}) reduces to the Schwarzschild solution. To set the stage, consider a spacetime in which quintessence, characterized by its density $\rho_q$, is present with
\begin{equation}
    \rho_q=-\frac{p}{2} \frac{3\omega}{r^{3(1+\omega)}}.
\end{equation}
 We know that in a commutative spacetime, one has a point mass described by the Dirac delta; however, noncommutative geometry naturally smears point sources over a finite length scale $\sqrt{\Theta}$, replacing the Dirac delta. The non-commutative inspired black hole was first constructed using a Gaussian mass profile \citep{Nicolini:2005vd}, and alternative smeared densities have since been proposed (see e.g. \citep{Giri:2006rc, Anacleto:2019tdj, Hamil:2024ppj}). In this work, we adopt a regular mass distribution $\rho_\Theta(r)$ so that the usual mass parameter $M$ in Eq.~\eqref{fr} is replaced by a radially dependent smeared mass $M(r,\Theta)$. This procedure regularises the central singularity at scales $r\lesssim\sqrt{\Theta}$ and recovers the Kiselev/non-commutative limits when $\Theta\to0$ and $p\to0$, respectively.

 By introducing a minimal length scale through the smearing of matter distributions, the model inspired by noncommutative geometry sheds light on the nature of gravity. In this framework, spacetime coordinates are treated as noncommuting operators satisfying the relation
\begin{equation}
[x_a, x_b] = i \theta_{ab},
\end{equation}
where $\theta_{ab}$ is a $4 \times 4$ antisymmetric matrix that defines the fundamental discretisation scale of spacetime. This $4 \times 4$ antisymmetric matrix has $6$ independent components  (similar to an electromagnetic field tensor). To replace it by a single scalar noncommutativity scale, we need a Lorentz-invariant quantity. The natural invariant of an antisymmetric 2-form is defined as $\theta_{ab} \theta^{ab}$, which is represented by $\Theta$ in this paper.
This noncommutativity leads to a generalised uncertainty principle (GUP)
\begin{equation}\label{up}
\Delta x^{\mu}\, \Delta x^{\nu} \ge \frac{1}{2}|\theta^{\mu\nu}|,
\end{equation}
indicating that spacetime points cannot be localized with arbitrary precision. This framework refines semiclassical gravity by incorporating noncommutative effects that are expected to appear in a quantum theory of gravity. Nicolini \textit{et al.} \cite{Nicolini:2005vd} first realized this idea by constructing a noncommutative geometry--inspired Schwarzschild black hole as an exact solution of Einstein's equations with a static, spherically symmetric, Gaussian-smeared matter source.

 The uncertainty relation indicates that spacetime cannot be sharply defined, leading to an intrinsic fuzziness Hamil~\citep{Hamil:2024ppj}. The noncommutative parameter $\Theta$, a small positive constant, measures this fuzziness and sets the scale of the minimal length. Different smeared mass distributions have been discussed in the literature~\citep{Giri:2006rc, Anacleto:2019tdj, Nicolini:2005gy, Kumar:2017hgs, Ghosh:2017odw, Ghosh:2020cob, AraujoFilho:2025rwr, AraujoFilho:2025jcu, Filho:2024zxx}. In the present work, we follow the form proposed by Anacleto \textit{et al.}~\citep{Anacleto:2019tdj}:
\begin{equation}
\rho_{\Theta}(r)=\frac{M\sqrt{\Theta}}{r^{3/2}(r^2+\pi\Theta)^2}.
\end{equation}
The corresponding mass function is obtained as
\begin{eqnarray}
M(r,\Theta)&=&4\pi\int^r r^2\rho_{\Theta}(r)\,dr \nonumber\\
&=&\frac{2M}{\pi}\left[\tan^{-1}\!\left(\frac{r}{\sqrt{\pi\Theta}}\right)-\frac{r\sqrt{\pi\Theta}}{r^2+\pi\Theta}\right].
\end{eqnarray}
For small values of $\Theta$, the first-order correction to the mass function can be written as
\begin{equation}
M(r,\Theta)\approx M-\frac{4M}{\sqrt{\pi}r}\sqrt{\Theta}.
\end{equation}
Hence, in a noncommutative spacetime, the mass of a particle is not concentrated at a point but spread over a region of size $\sqrt{\Theta}$. Using this smeared mass distribution, the Kiselev black hole metric becomes
\begin{equation}
f(r,\Theta)=1-\frac{2M}{r}+\frac{8M}{\sqrt{\pi}r^2}\sqrt{\Theta}-\frac{p}{r^{3\omega+1}}.
\end{equation}
In the limiting case $\Theta \to 0$, the spacetime reduces to the Kiselev black hole surrounded by quintessence~\citep{Kiselev:2002dx}, while for $p \to 0$, it approaches the noncommutative Schwarzschild solution~\citep{Nicolini:2005vd}. When both $\Theta \to 0$ and $p \to 0$, the standard Schwarzschild geometry is recovered.

\begin{figure*}
\begin{tabular}{c c c}
\includegraphics[scale=0.55]{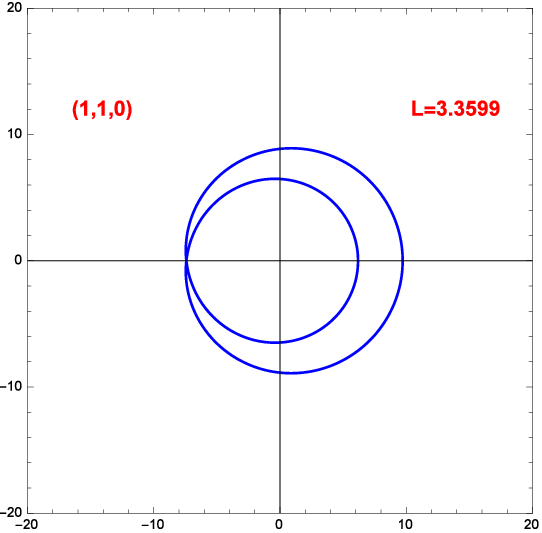}&
\includegraphics[scale=0.55]{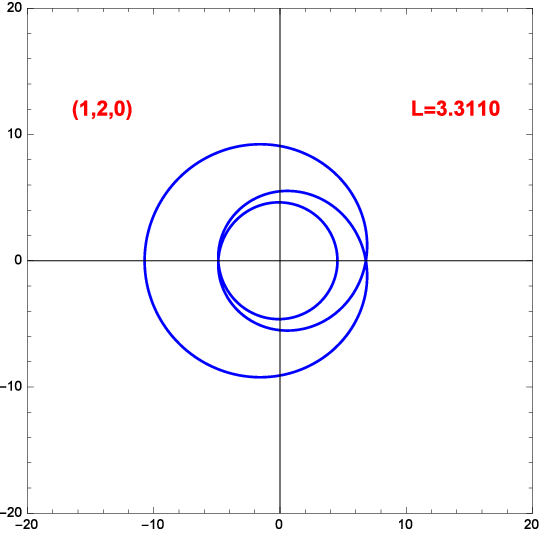}&
\includegraphics[scale=0.55]{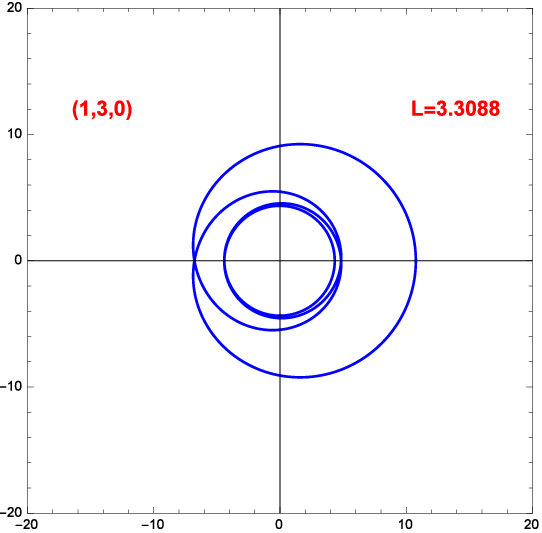}\\
\includegraphics[scale=0.55]{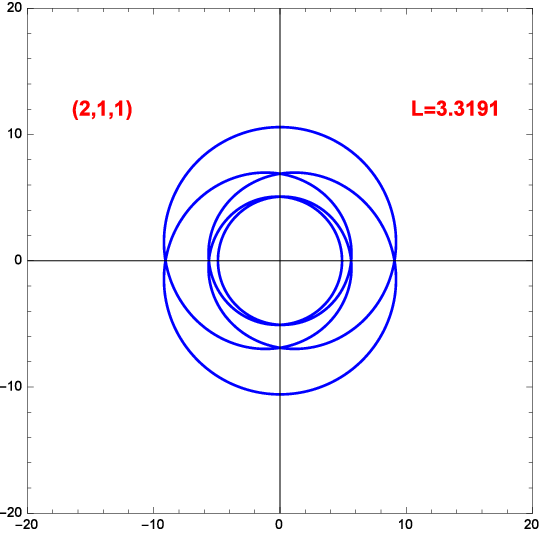}&
\includegraphics[scale=0.55]{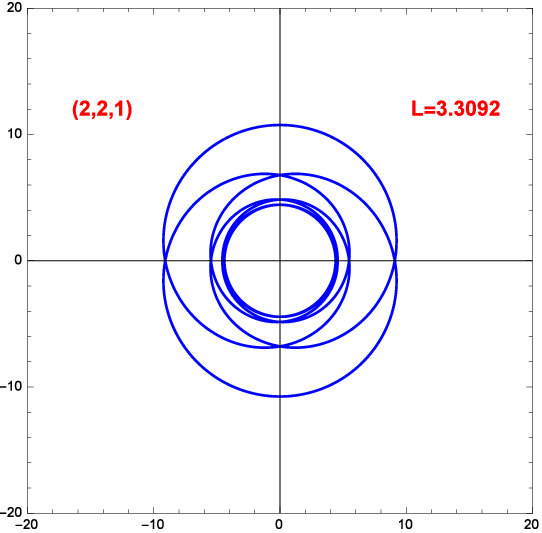}&
\includegraphics[scale=0.55]{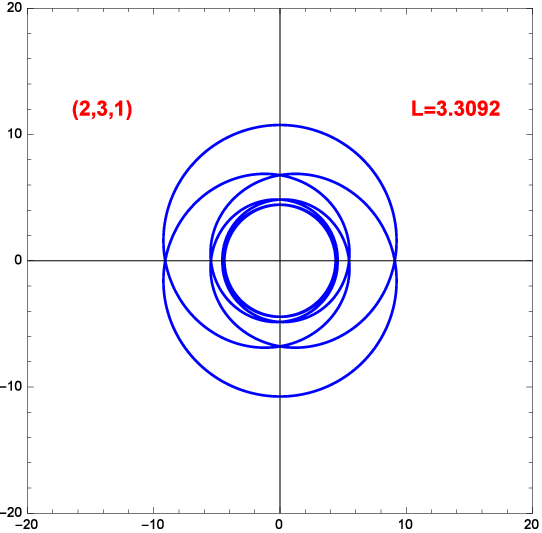}\\
\includegraphics[scale=0.55]{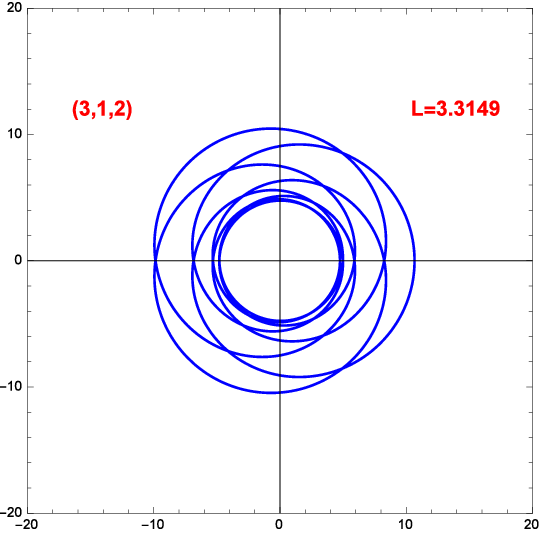}&
\includegraphics[scale=0.55]{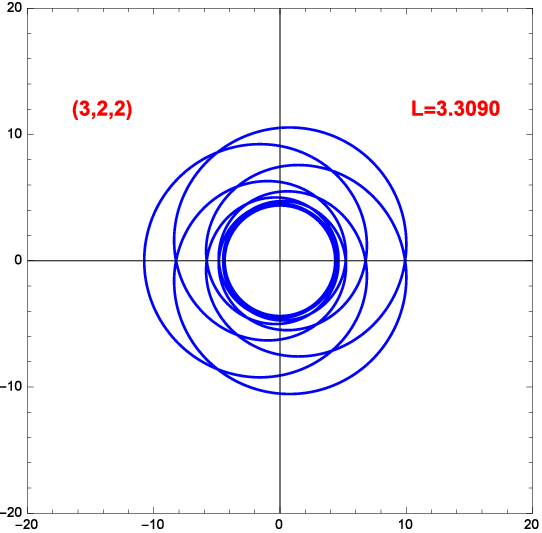}&
\includegraphics[scale=0.55]{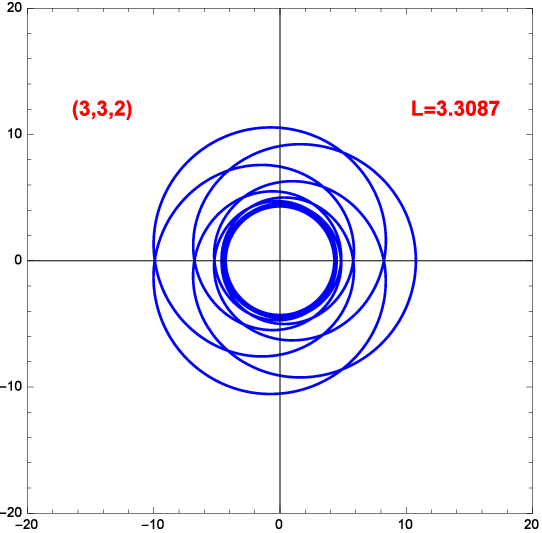}\\
\includegraphics[scale=0.55]{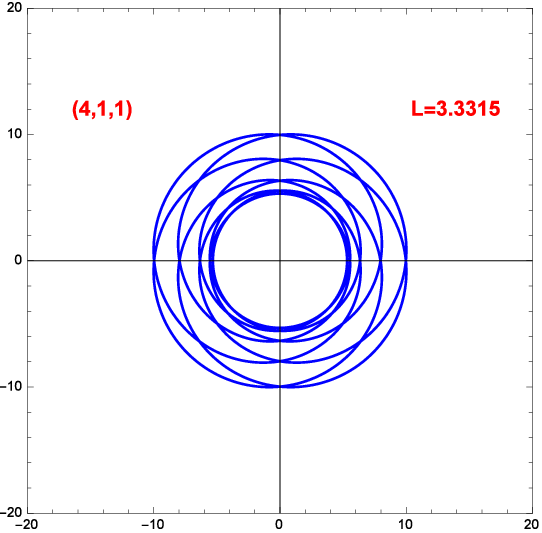}&
\includegraphics[scale=0.55]{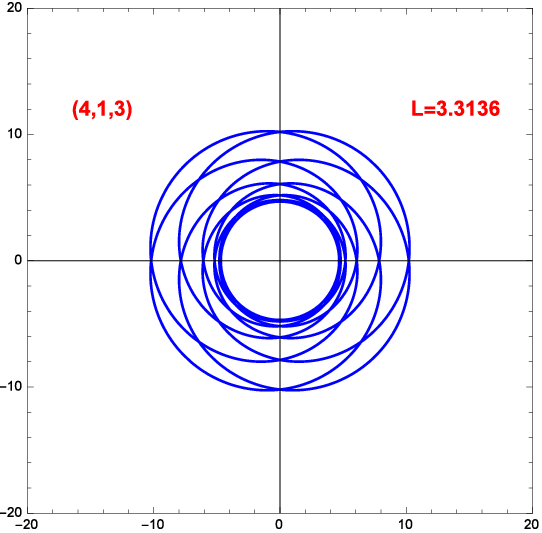}&
\includegraphics[scale=0.55]{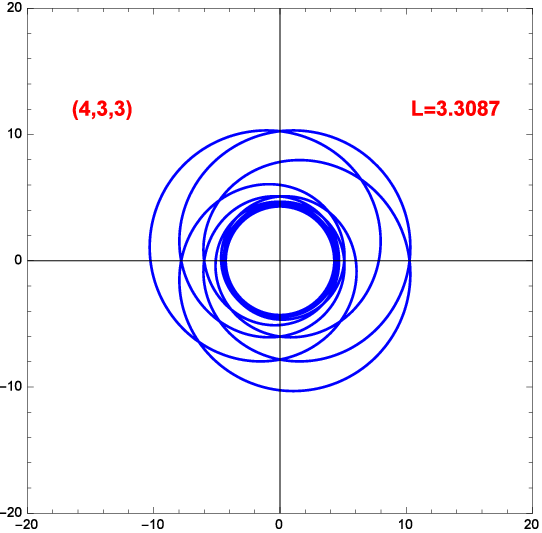}\\
\end{tabular}	
\caption{Periodic orbits around a non-commutative-inspired black hole surrounded by quintessence with an equation-of-state parameter $\omega = -2/3$. The non-commutative parameter is set to $\Theta = 0.01$ and the particle energy to $E = 0.94$. Each trajectory corresponds to a different set of zoom–whirl–vertex numbers $(z, w, v)$, illustrating the geometric complexity and structure of the bound periodic orbits.}
\label{per-orb1} 
\end{figure*}

\section{Periodic orbits}\label{section3}
\renewcommand{\theequation}{3.\arabic{equation}} \setcounter{equation}{0}

The periodic time-like orbits around a black hole enveloped by quintessence and inspired by noncommutative geometry are covered in this section. Understanding the complex structure of bound trajectories in strong gravitational fields requires an examination of periodic orbits ~\cite{Levin:2008yp}.  Let us first consider the motion of a test particle in the spacetime of a black hole. The Lagrangian that governs the dynamics of the particle reads 
\bqn
{\cal L} = \frac{1}{2} g_{\mu\nu} \frac{dx^\mu}{d\tau} \frac{dx^\nu}{d\tau},
\eqn
where $\tau$ denotes the proper time, which serves as the affine parameter along the world line of a timelike particle. For a massless particle, ${\cal L} = 0$, while for a massive one ${\cal L} < 0$.  

The corresponding generalized momentum $p_\mu$ is given by
\bqn
p_\mu = \frac{\partial {\cal L}}{\partial \dot{x}^\mu} = g_{\mu\nu} \dot{x}^\nu,
\eqn
which leads to the following conserved quantities for a stationary and axisymmetric spacetime:
\bqn
p_t &=& g_{tt} \dot{t} = -E,\\
p_\phi &=& g_{\phi\phi} \dot{\phi} = L_z,\\
p_r &=& g_{rr} \dot{r},\\
p_\theta &=& g_{\theta\theta} \dot{\theta},
\eqn
where $E$ and $L_z$ represent, respectively, the conserved energy and angular momentum per unit mass of the particle. A dot denotes differentiation with respect to the affine parameter $\lambda$.  

From these definitions, we obtain
\bqn\lb{dot1}
\dot{t} = -\frac{E}{g_{tt}} = \frac{E}{f(r,\Theta)},\\
\lb{dot2}
\dot{\phi} = \frac{L_z}{g_{\phi\phi}} = \frac{L_z}{r^2 \sin^2\theta}.
\eqn
For timelike geodesics, the normalization condition
\begin{equation}
g_{\mu\nu} \dot{x}^\mu \dot{x}^\nu = -1
\end{equation}
must hold. Substituting Eqs.~(\ref{dot1}) and (\ref{dot2}) into this relation yields
\bqn
g_{rr}\dot{r}^2 + g_{\theta\theta}\dot{\theta}^2 &=& -1 - g_{tt}\dot{t}^2 - g_{\phi\phi}\dot{\phi}^2 \nb\\
&=& -1 + \frac{E^2}{f(r,\Theta)} - \frac{L_z^2}{r^2 \sin^2\theta},
\eqn
which governs the radial and polar motion of a test particle in the  noncommutative geometry. The study of such orbits provides a natural framework to classify zoom–whirl periodic trajectories and to explore their observational signatures in noncommutative black hole spacetimes~\cite{Levin:2008yp,Levin:2009sk,Misra:2010pu}.

We are interested in the evolution of particles in equatorial circular orbits. For simplicity, we choose $\theta=\pi/2$ and $\dot \theta=0$. Then the above expression can be simplified into the form
\bqn\lb{rdot}
\dot r ^2 = E^2 - V_{\rm eff}(r),
\eqn
where $V_{\rm eff}(r)$ denotes the effective potential and is given by
\bqn \lb{Veff}
V_{\rm eff}(r)= \left(1+\frac{L_z^2}{r^2}\right)f(r,\Theta).
\eqn
One immediately observes that $V_{\rm eff}(r) \to 1$ as $r \to +\infty$, as expected for an asymptotically flat spacetime. In this case, particles with energy $E >1$ can escape to infinity. The case $E = 1$ is the critical point between bound and unbound orbits. Thus, the maximum energy for the bound orbits is $E=1$. We can obtain the trajectory of a particle by integrating Eqs.~(\ref{dot1}), (\ref{dot2}), and (\ref{Veff}) to get $t$, $\phi$, and $r$ as functions of $\tau$. However, since Eq.~(\ref{rdot}) involves taking a square root, the choice of sign corresponds to whether the particle is moving inward or outward, and must be specified manually before any numerical integration. A convenient equation of motion, derived from the $r-$component of the geodesic equation, can be used for numerical analysis:
\bqn
\ddot{r}=\frac{f'(r,\Theta)}{2f(r,\Theta)}\dot{r}^2-\frac{f'(r,\Theta)E^2}{2f(r,\Theta)}+\frac{f(r,\Theta)L_z^2}{r^3}.
\eqn
This equation is convenient for numerical integration and helps in understanding the stability of circular orbits, as well as how they evolve into periodic or zoom-whirl trajectories in strong gravitational fields~\cite{Levin:2008yp,Levin:2009sk,Misra:2010pu,Chandrasekhar:1985kt}.

\begin{figure*}
\begin{tabular}{c c c}
\includegraphics[scale=0.55]{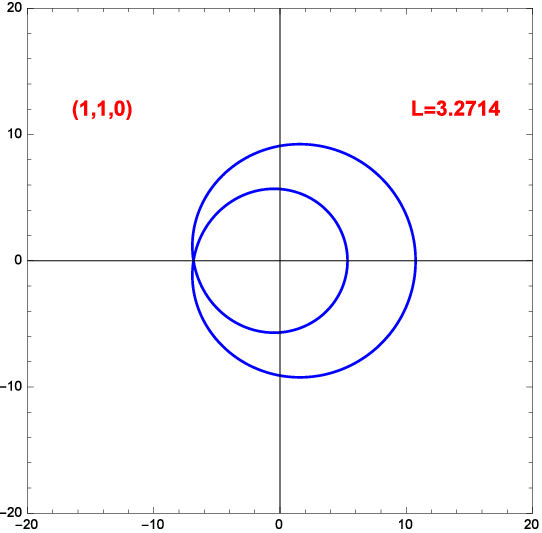}&
\includegraphics[scale=0.55]{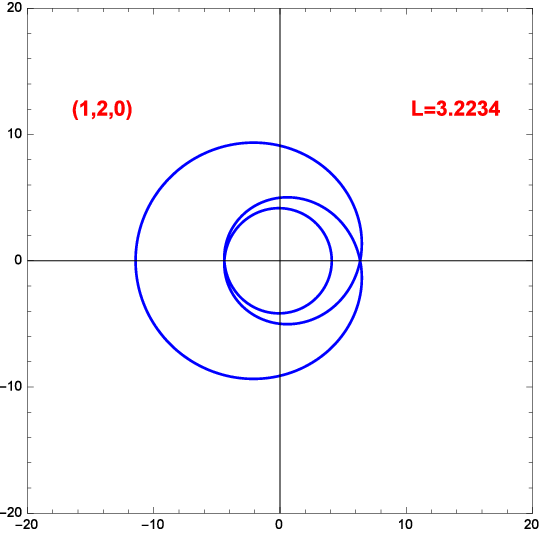}&
\includegraphics[scale=0.55]{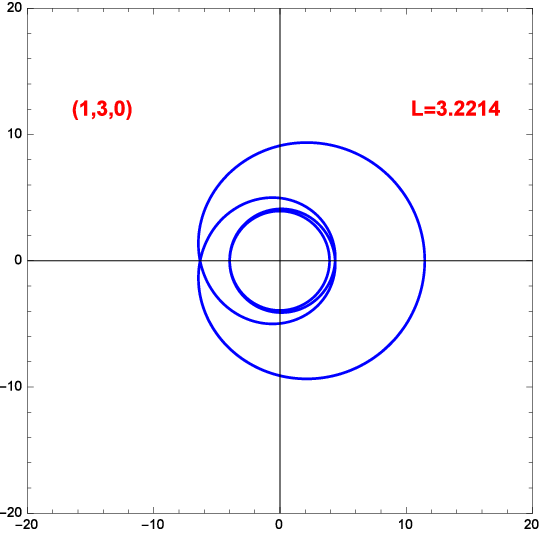}\\
\includegraphics[scale=0.55]{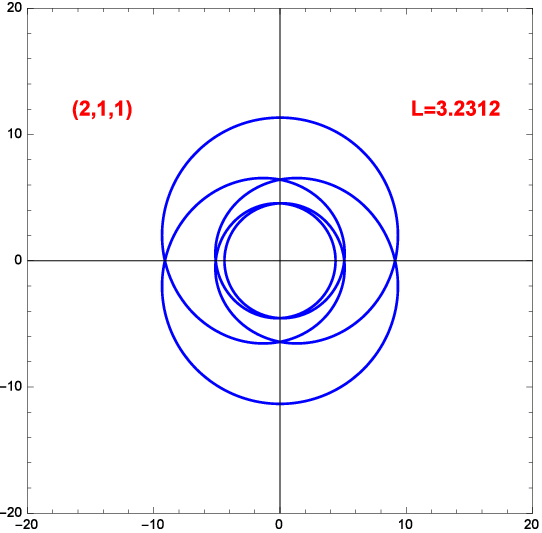}&
\includegraphics[scale=0.55]{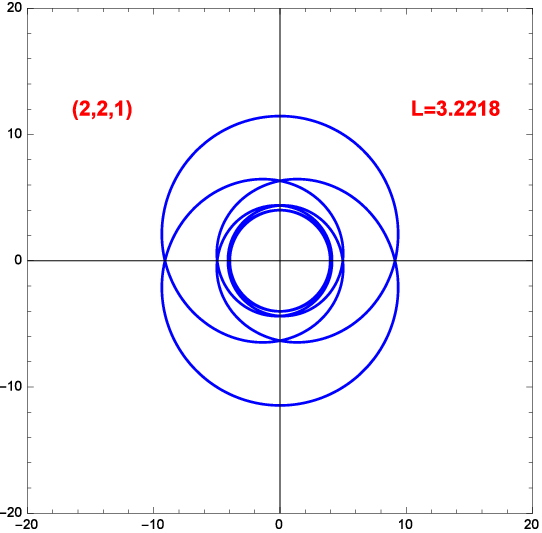}&
\includegraphics[scale=0.55]{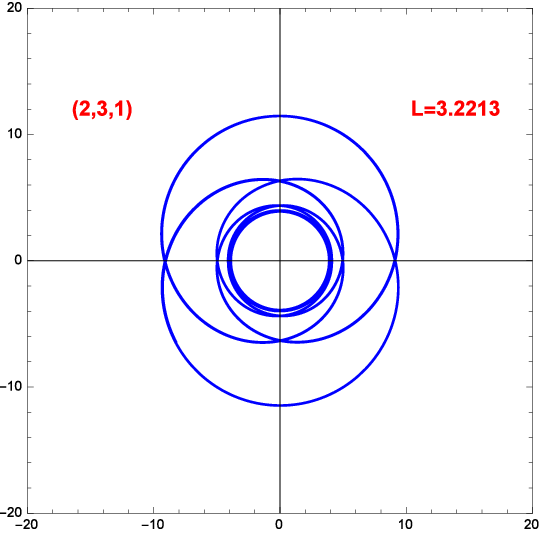}\\
\includegraphics[scale=0.55]{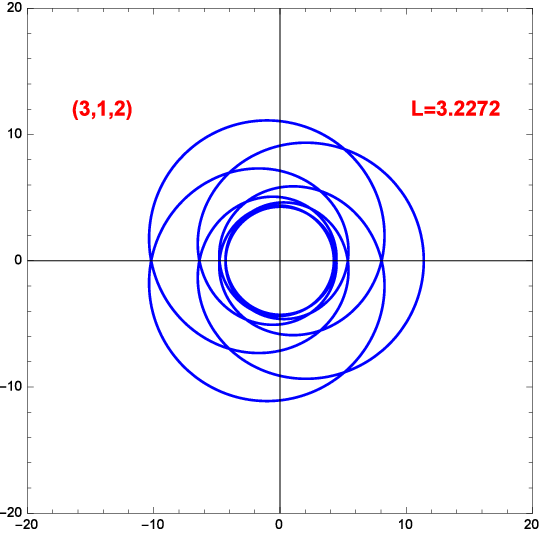}&
\includegraphics[scale=0.55]{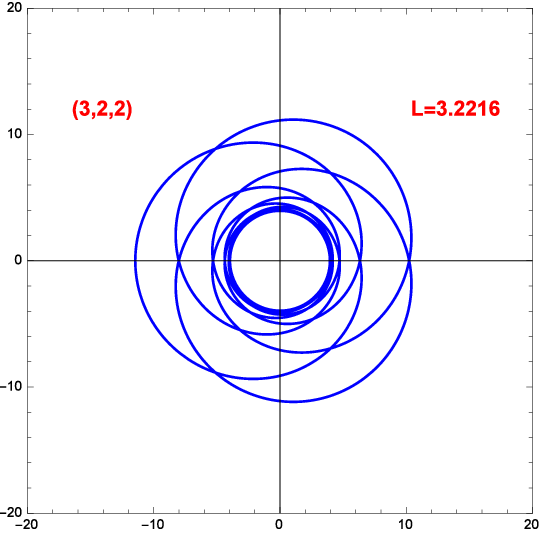}&
\includegraphics[scale=0.55]{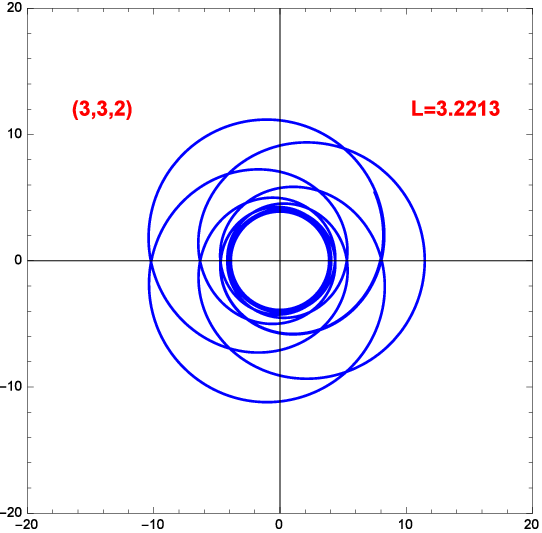}\\
\includegraphics[scale=0.55]{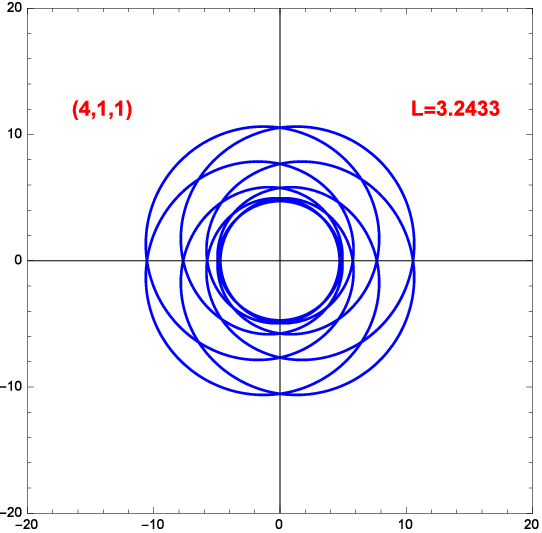}&
\includegraphics[scale=0.55]{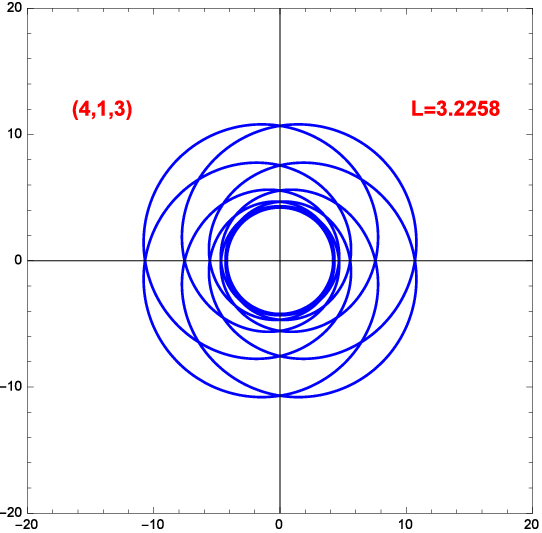}&
\includegraphics[scale=0.55]{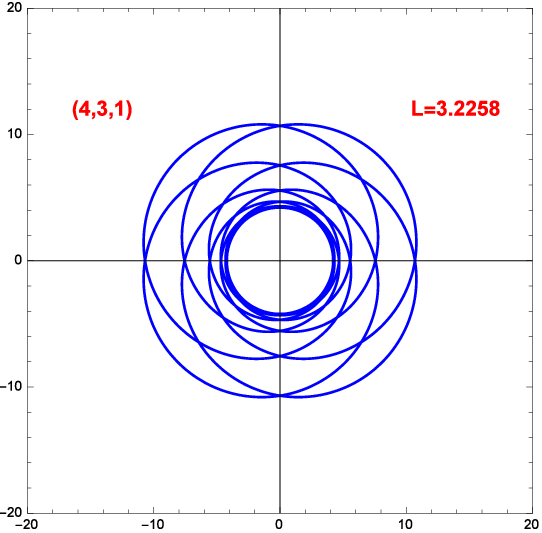}\\
\end{tabular}	
\caption{Periodic orbits for various $(z, w, v)$ combinations around a non-commutative-inspired black hole surrounded by quintessence with $\omega = -2/3$. Here, the non-commutative parameter is increased to $\Theta = 0.02$ while keeping the particle energy fixed at $E = 0.94$. Increasing $\Theta$ slightly modifies the orbit shape, leading to broader zoom regions and altered precession characteristics.}

\label{per-orb2} 
\end{figure*}

After the integration is complete, a periodic orbit can be obtained for given values of $E$ and $L_z$. A periodic orbit is a bound trajectory that returns exactly to its initial position after a fixed period. Such orbits can take various shapes, depending on the particle's energy and angular momentum. To study them systematically, it is convenient to employ a classification scheme.

We adopt the recipe introduced by Levin and Perez-Giz~\cite{Levin:2008mq}, which classifies all periodic orbits around black holes using a triplet of integers $(z, w, v)$, corresponding to the zoom, whirl, and vertex behavior of the trajectory. In their scheme, a periodic orbit returns to its initial conditions after a finite time, which requires that the ratio of radial to azimuthal frequencies be a rational number. Because a rational one can approximate any irrational number, periodic orbits can effectively represent generic bound trajectories around black holes. The Levin and Perez-Giz~\cite{Levin:2008mq} recipe has been successfully applied to various black holes, including Schwarzschild and Kerr geometries~\cite{Levin:2009sk, Misra:2010pu, Babar:2017gsg}, and provides a useful framework for studying the corresponding gravitational radiation from these orbits.

According to the taxonomy of~\cite{Levin:2008mq}, the ratio $q$ between the two frequencies $\omega_r$ and $\omega_\phi$ of oscillations in the $r$-motion and $\phi$-motion, respectively, in terms of three integers $(z, w, v)$ as
\bqn
q \equiv \frac{\omega_\phi}{\omega_r}-1 = w + \frac{v}{z}.
\eqn
The integers $(z, w, v)$ each have different geometric meanings. The zoom number $z$ counts the larger circles in an orbit, while the whirl number $w$ counts the small loops near the center. The vertex number $v$ tells us if the particle moves through the orbit's vertices in a clockwise or counterclockwise direction. To avoid degeneracy, $z$ and $v$ should be relatively prime~\cite{Levin:2008mq}. The parameter $q$ indicates the degree to which the periapsis deviates from that of a simple ellipse, allowing us to understand the orbit's shape. This framework also considers the order in which the orbital paths or segments are traced. Together, all these numbers help describe the complex behavior of periodic orbits. The ratio $\frac{\omega_\phi}{\omega_r}$ equals $\Delta \phi/(2\pi)$, where $\Delta \phi = \oint d\phi$ is the total equatorial angle during a period in $r$, and this must be a multiple of the total number of $2\pi$. With the geodesic equations for non-commutative black holes, we can calculate $q$ as follows:
\bqn
q &=& \frac{1}{\pi} \int_{r_1}^{r_2} \frac{\dot \phi}{\dot r} dr -1\nb\\
&=& \frac{1}{\pi} \int_{r_2}^{r_1} \frac{L_z}{\sqrt{E^2- V_{\rm eff}(r)}}dr-1,
\eqn
where $r_1$ and $r_2$ are two turning points. 

The behavior of $q$ as $E$ and $L_z$ vary can be found in~\cite{Azreg-Ainou:2020bfl}. In Figs.~\ref{per-orb1} and \ref{per-orb2}, we illustrate the periodic orbits of non-commutative black holes for different combinations of integers $(z, w, v)$. It is worth mentioning that we set $M=1$ for simplicity in the figures. In addition, we fixed $\omega=-2/3$ and $p=0.001$ for all calculations. The value of $z$ determines the number of blades in the orbit’s shape. The larger $z$ values correspond to larger blade profiles and increasingly complex trajectories. 

\section{Gravitational Radiation in non-commutative geometry}\label{section4}
\renewcommand{\theequation}{4.\arabic{equation}} \setcounter{equation}{0}

In this section, we present a preliminary analysis of the gravitational radiation emitted by a test particle moving in periodic orbits around SMBHs modelled by our noncommutative geometry–inspired solution. The Extreme Mass Ratio Inspirals (EMRIs), consisting of a stellar-mass compact object orbiting an SMBH, are among the most promising sources for future space-based GW detectors such as LISA, Taiji, and TianQin~\cite{LISA:2017pwj,Gong:2021gvw,Hu:2017mde}. The GWs generated by these systems encode detailed information about the strong-field dynamics and the underlying spacetime geometry of the central black hole. A possible observational test of quantum gravity-inspired models could be provided if the smaller body travels on a periodic orbit in a noncommutative spacetime, where spacetime coordinates obey nontrivial commutation relations, and its emitted waveform carries imprints of noncommutative effects. 

The analysis of gravitational waveforms from EMRIs is typically carried out using the adiabatic approximation, which assumes that the inspiral timescale is much longer than the orbital period~\cite{Hughes:1999bq,Barack:2003fp}. The motion of the smaller object can be described as a series of geodesics in the SMBH's background metric, as its energy and angular momentum change slowly in this regime. For short-term orbital evolution~\cite{Isoyama:2021jjd}, the radiation response, or back-reaction of the emitted GWs on the particle's motion, is ignored at leading order.

\begin{figure*}
\begin{tabular}{c c}
\includegraphics[scale=0.9]{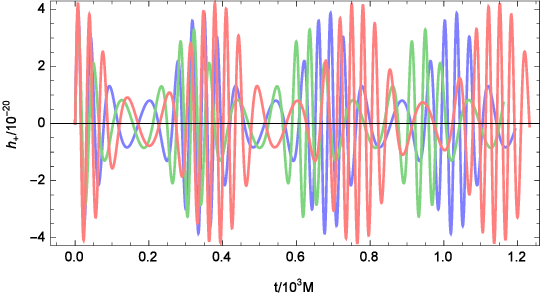}&
\includegraphics[scale=0.9]{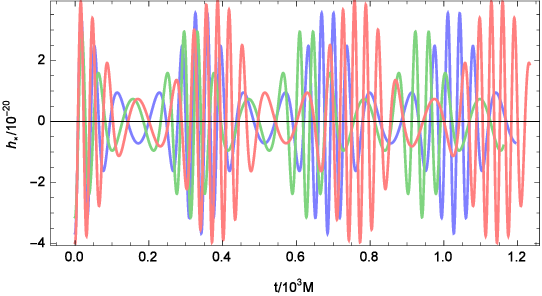}\\
\end{tabular}
\caption{Gravitational waveforms (plus and cross polarizations) generated by a test particle of mass $m = 10M_{\odot}$ in periodic orbits characterized by $(z, w, v) = (1,2,0)$ (blue), $(2,1,1)$ (green), and $(3,2,2)$ (red) around a supermassive black hole of mass $M = 10^{7} M_{\odot}$. The non-commutative parameter is $\Theta = 0.01$ and $E = 0.94$. Distinct zoom–whirl phases in the orbital motion are reflected in the modulation of the waveform amplitude and frequency.}

\label{gwpolar1} 
\end{figure*}

We employ a waveform model that offers a practical framework for computing the GWs emitted by periodic orbits in a black hole spacetime, following the approach developed in~\cite{Babak:2006uv} -- often referred to as the numerical kludge scheme-- that proceeds in two main steps. First, the motion of the small compact object is obtained by numerically integrating the geodesic equations in the background spacetime of the black hole. In the second step, the corresponding gravitational waveform is constructed using the standard quadrupole formula for gravitational radiation. This semi-relativistic approximation has been widely used to model GW signals from EMRIs and provides a powerful tool for analyzing the dynamics of the orbit, the properties of the central black hole, and possible environmental effects~\cite{Barack:2003fp, Gair:2004iv, Hughes:2000ssa}.
For a metric perturbation $h_{ij}$ representing the GW and a symmetric, trace-free (STF) mass quadrupole moment $I_{ij}$, the quadrupole formula takes the form
\begin{equation}
h_{ij}=\frac{1}{A}\ddot{I}_{ij},
\end{equation}
where $A=c^4 D_L/(2G)$, $G=c=1$, and $D_L$ is the luminosity distance to the source. By numerically solving the geodesic equations, one obtains the trajectory $Z_i(t)$ of the small object in the curved spacetime of the supermassive black hole, which is then used to compute the GW signal. For a particle of mass $m$ moving along a trajectory $Z^i(t)$, the quadrupole moment $I_{ij}$ is defined as~\cite{Thorne:1980ru}
\begin{equation}\label{lvalue}
I^{ij}=m\int d^3x\, x^i x^j\, \delta^3(x^i - Z^i(t)).
\end{equation}

\begin{figure*}
\begin{tabular}{c c}
\includegraphics[scale=0.9]{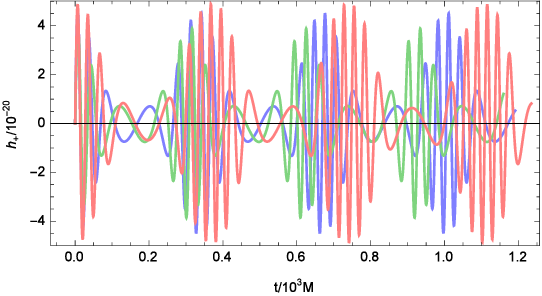}&
\includegraphics[scale=0.9]{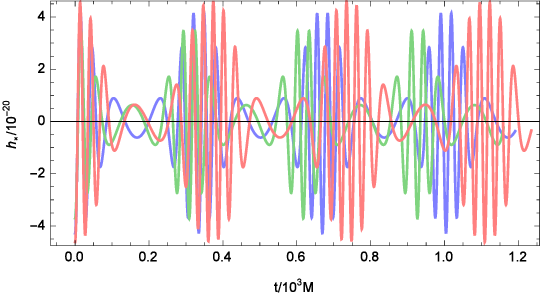}\\
\end{tabular}
\caption{Gravitational waveforms from a test object with $m=10 M_\odot$ around periodic orbits $(1,2,0)$: blue, $(2,1,1)$: green, and $(3,2,2)$: red, around a supermassive black hole with mass $M=10^7 M_\odot$. The value of parameter $\Theta=0.02$ and energy is fixed at $E=0.94$. The left and right panels correspond to plus and cross polarizations, respectively.} 
\label{gwpolar2} 
\end{figure*}

The choice of coordinate system plays a key role in both the computation and interpretation of gravitational waveforms. While the geodesic equations are usually solved in the coordinates $(r, \theta, \phi)$, the resulting waveform is conveniently expressed in a detector-adapted Cartesian coordinates $(X, Y, Z)$, which simplifies the analysis of the signal measured by a gravitational-wave detector. The transformation   is given by~\cite{Babak:2006uv}
\begin{equation}\label{4.3}
x = r \sin\theta \cos\phi, \quad 
y = r \sin\theta \sin\phi, \quad 
z = r \cos\theta.
\end{equation}
This transformation enables us to project the trajectory of the small object onto a Cartesian grid, which is necessary for evaluating the source multipole moments. The metric perturbations $h_{ij}$, representing the emitted gravitational waves, are then calculated from the second time derivative of the mass quadrupole moment $I_{ij}$ as
\begin{equation}\label{4.4}
h_{ij} = \frac{m}{A}\left(a_i x_j + a_j x_i + 2 v_i v_j\right),
\end{equation}
where $v_i$ and $a_i$ denote the velocity and acceleration components of the small object, respectively, and $A = c^4 D_L / (2G)$ with $G = c = 1$. 
This formalism adheres to the conventional method of numerical kludge waveforms~\cite{Barack:2003fp, Gair:2004iv, Babak:2006uv, Hughes:2000ssa}, which provides an effective and physically consistent approach to approximating EMRI waveforms.

To analyze the gravitational-wave signal as observed by a detector, it is convenient to introduce a detector-adapted Cartesian coordinate system $(X, Y, Z)$, centred on the black hole and oriented with respect to the source frame $(x, y, z)$ by the inclination angle $\iota$ and the longitude of pericentre $\zeta$~\cite{Babak:2006uv, Barack:2003fp, Gair:2004iv}. This transformation facilitates the projection of the waveform onto the detector frame, enabling the computation of the observable GW polarisations.
 The unit vectors of the detector frame in the $(x, y, z)$ coordinates are:
\begin{eqnarray}
\hat{e}_X &=& (\cos\zeta, -\sin\zeta, 0),\\
\hat{e}_Y &=& (\sin\iota \sin\zeta, \cos\iota \cos\zeta, -\sin\iota),\\
\hat{e}_Z &=& (\sin\iota \sin\zeta, -\sin\iota \cos\zeta, \cos\iota),
\end{eqnarray}
The GW polarizations $h_+$ and $h_\times$ are then obtained by projecting $h_{ij}$, Eq. (\ref{4.4}), onto the detector frame
\begin{eqnarray}\label{4.5}
h_+&=\frac{1}{2}\big(e_X^i e_X^j-e_Y^ie_Y^j\big)h_{ij},\\
h_{\times}&=\frac{1}{2}\big(e_X^i e_Y^j-e_Y^ie_X^j\big)h_{ij},
\end{eqnarray}
These polarizations can be expressed in terms of components $h_{\zeta\zeta}$, $h_{\iota\iota}$, and $h_{\iota\zeta}$, which are defined in the detector frame as combinations of the $h_{ij}$ components as
\begin{eqnarray}\label{4.64}
h_+&=&\frac{1}{2}\big(h_{\zeta\zeta}-h_{\iota\iota}\big),\\\label{4.6}
h_{\times}&=&h_{\iota\zeta},
\end{eqnarray}
where the components are \cite{Babak:2006uv}
\begin{eqnarray}\label{4.7}
h_{\zeta\zeta}&=&h_{xx}\cos^2\zeta-h_{xy}\sin{2 \zeta}+h_{yy}\sin^2\zeta,\\
h_{\iota\iota}&=& \cos^2\iota\big[h_{xx}\sin^2\zeta + h_{xy}\sin 2 \zeta + h_{yy}\cos^2 \zeta\big] \nonumber \\
&& +h_{zz} \sin^2\iota - \sin{2 \iota}\big[h_{xz \sin\zeta}+h_{yz}\cos\zeta\big],\\
h_{\iota\zeta}&=& \frac{1}{2}\cos\iota\big[h_{xx} \sin {2 \zeta}+ 2 h_{xy}\cos{2 \zeta}- h_{yy}\sin{2 \zeta}\big]\nonumber\\
&&+\sin\iota\big[h_{yz} \sin\zeta-h_{xx} \cos\zeta\big].
\end{eqnarray}

\begin{figure*}
\begin{tabular}{c c}
\includegraphics[scale=0.95]{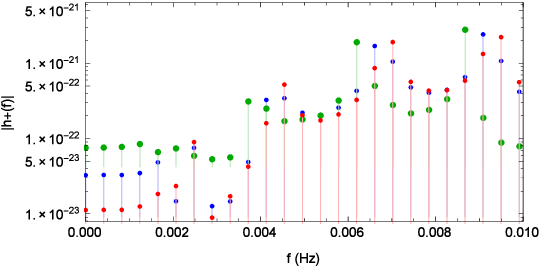}& 
\includegraphics[scale=0.95]{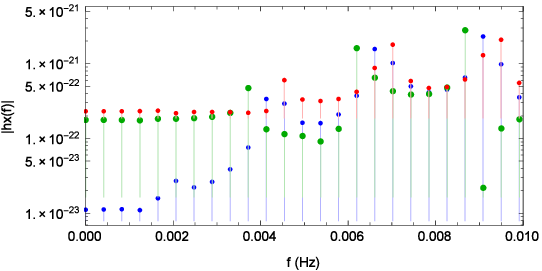}\\
\end{tabular}
\caption{Fourier spectra $|\tilde{h}_{+,\times}(f)|$ corresponding to the time-domain waveforms shown in Fig.~\ref{gwpolar1} for $\Theta = 0.01$. The spectral peaks correspond to characteristic frequencies of the zoom–whirl orbits, showing distinct harmonic structures related to the orbital parameters $(z, w, v)$.}

\label{freq-spect1} 
\end{figure*}

\begin{figure*}
\begin{tabular}{c c}
\includegraphics[scale=0.95]{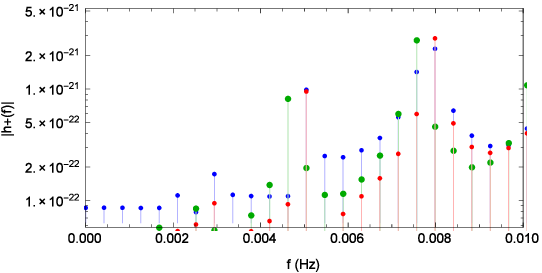}& 
\includegraphics[scale=0.95]{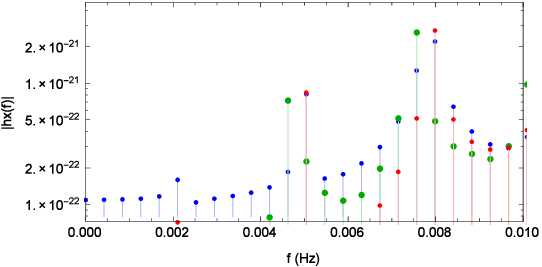}\\
\end{tabular}
\caption{Fourier spectra $|\tilde{h}_{+,\times}(f)|$ for the gravitational waveforms in Fig.~\ref{gwpolar2} with $\Theta = 0.02$. Increasing the non-commutative parameter shifts the spectral peaks and enhances the high-frequency components, indicating stronger gravitational radiation and modified orbital dynamics.}

\label{freq-spect2} 
\end{figure*}

To examine the influence of the noncommutative parameter on gravitational waveforms generated by different periodic orbits in an EMRI system, we consider a compact object of mass $m = 10\,M_\odot$ orbiting a supermassive black hole (SMBH) of mass $M = 10^7\,M_\odot$. For simplicity, the inclination angle $\iota$ and the longitude of pericentre $\zeta$ are fixed at $\pi/4$, and a luminosity distance of $D_L = 2$~Gpc is adopted for the computation of the GW polarisations. 

The resulting gravitational waveforms, represented by the two independent components $h_+$ and $h_\times$, exhibit a characteristic alternating pattern. During the portions of the orbit where the trajectory extends outward in a highly eccentric fashion (the zoom phases), the waveform amplitude remains relatively low. These intervals are followed by short, intense bursts of radiation associated with the nearly circular segments of the trajectory (the whirl phases). The number of low-amplitude intervals corresponds to the number of zoom segments, while the number of intense bursts matches the number of whirls in the orbit. The numerical results obtained from Eqs.~(\ref{4.64}) and~(\ref{4.4}) are shown in Figs.~\ref{gwpolar1} and~\ref{gwpolar2}, which clearly display the distinct ``zoom'' and ``whirl'' features of the GW signal from periodic orbits in EMRIs, reflecting the orbital dynamics of the small object over one complete cycle~\cite{Levin:2008mq, Barack:2003fp, Babak:2006uv, Gair:2004iv}.

In Fig.~\ref{gwpolar1}, the gravitational waveforms are shown with $(z,w,v)=(1,2,0),(2,1,1)$ and $(3,2,2)$. This analysis reveals a strong correlation between gravitational waveforms and the orbital motion of the small object. Each orbit displays clear ``zoom" and ``whirl" phases in the waveform that mirror the corresponding behaviors in the object's trajectory.
%Additionally, orbits with higher zoom numbers $z$ produce waveforms with more intricate substructures, reflecting the increased number of ``leaves" in the full periodic orbit.

The presence of the noncommutative parameter $\Theta$ has a pronounced effect on the gravitational waveform generated by a massive particle moving in a periodic orbit. We again consider the periodic orbit shown in Fig.~\ref{gwpolar2} for different values of $\Theta$. Our study indicates that the gravitational waveforms exhibit a substantial change in amplitude and a discernible phase shift as $\Theta$ increases, demonstrating the impact of spacetime noncommutativity on the orbital dynamics and resultant radiation.

The gravitational waves emitted by a test particle in periodic motion around an SMBH  in a noncommutative spacetime can be further analyzed through their frequency spectra $\vert \tilde{h}_{+,\times}(f)\vert$ and characteristic strain $h_c(f)$, defined as
\begin{eqnarray}\label{ch}
h_c(f)=2f\left(\vert \tilde{h}_+(f)\vert^2+\vert \tilde{h}_\times(f)\vert^2\right)^{1/2}.
\end{eqnarray}
where, $\tilde h(f)=\int h(t)e^{-2\pi i f t}dt$ is the one-sided Fourier amplitude (positive frequencies only). This $h_c(f)$ is the intrinsic, polarization-combined strain (no detector antenna factors) and satisfies signal to noise ratio (SNR) as $\mathrm{SNR}^2=\int (h_c^2/ fS_n)d\ln f $ \cite{Depies:2009im, Larson:1999we}. 
The frequency spectra are obtained by applying a discrete Fourier transform (DFT) to the time-domain gravitational waveforms, converting the signal into the frequency domain. This transformation enables a detailed examination of the signal's frequency content, revealing how the particle’s periodic orbital motion modulates the structure of the emitted gravitational waves (see Figs.~\ref{freq-spect1} and~\ref{freq-spect2}). The dominant frequencies of these signals lie primarily in the millihertz range, making them especially relevant for space-based detectors such as LISA~\cite{LISA:2017pwj, Barack:2003fp, Gair:2004iv, Babak:2006uv}, which are designed to detect low-frequency gravitational waves from EMRIs. The characteristic spectra for different periodic orbits, labelled by the triplet $(z, w, v)$, are displayed in Figs.~\ref{freq-spect1} and~\ref{freq-spect2}. In the above analysis, Using a frequency resolution of $\approx 0.001/2 $ Hz and a maximum frequency of $0.01$ Hz (see Figs.~\ref{freq-spect1} and~\ref{freq-spect2}), the total duration of the signal is $ T \approx 2000 $s. For a sampling frequency of $0.01$ Hz, this corresponds to $N=20$ samples over the interval, yielding a time spacing of $ \Delta t \approx 100 $s per sample in Figs.~\ref{gwpolar1} and \ref{gwpolar2}.

\begin{figure*}
\begin{tabular}{c c}
\includegraphics[scale=0.55]{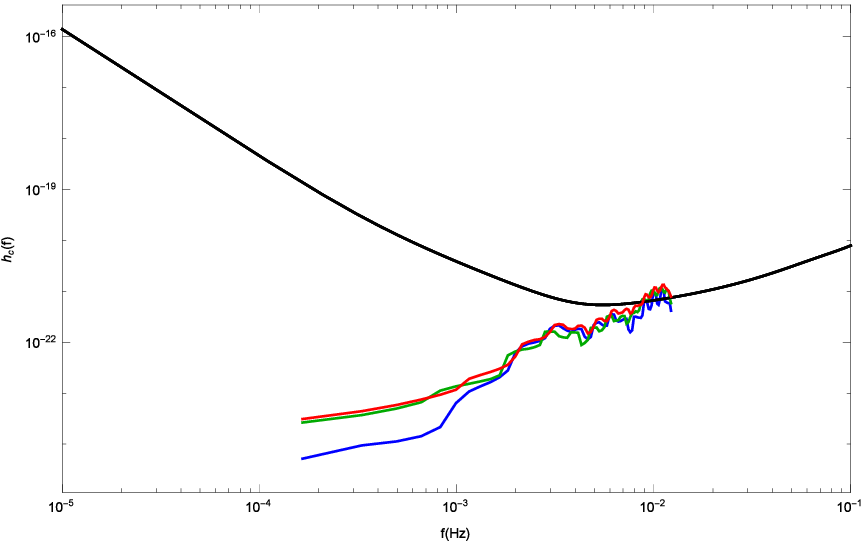}& 
\includegraphics[scale=0.55]{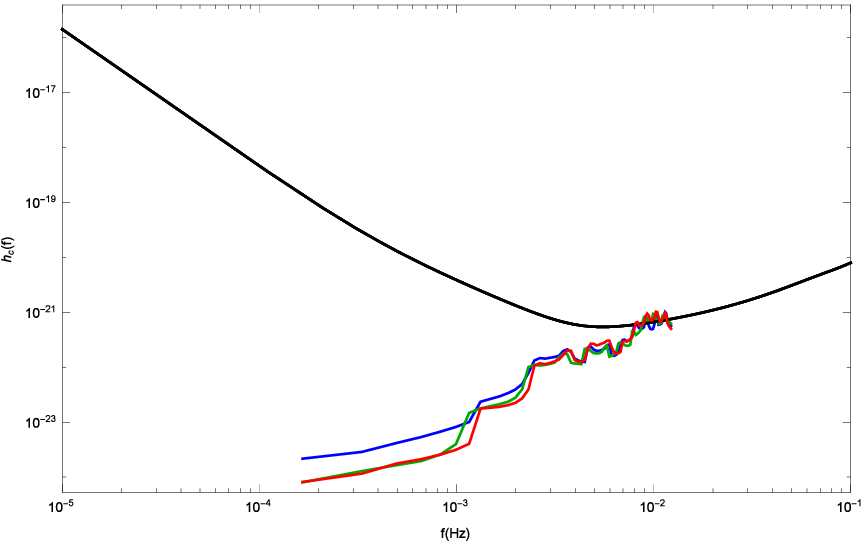}\\
\end{tabular}
\caption{Characteristics strain of gravitational waveforms of periodic orbits in Figs.~\ref{gwpolar1} (red) and \ref{gwpolar2} (right). The black curve corresponds to LISA sensitivity.  Portions of the spectra lie above the LISA sensitivity band, suggesting that these non-commutative effects could be detectable in future space-based gravitational wave observations.}
\label{strain} 
\end{figure*}

To enhance the visual clarity of the plots for the characteristic strain given in Eq.~(\ref{ch}), we applied a smoothing procedure to the numerically generated $h_c(f)$ by performing a running average over 30 frequency bins. Choosing a larger averaging window would further repress numerical noise but could also obscure fine spectral features. As shown in Fig.~\ref{strain}, portions of the characteristic strain corresponding to different orbital configurations $(z, w, v)$ and values of the noncommutative parameter $\Theta$ lie above the sensitivity curve of the Laser Interferometer Space Antenna (LISA). This means that the corresponding gravitational waves, exhibiting distinctive zoom–whirl features arising from spacetime noncommutativity, fall within the detectable range of LISA~\cite{LISA:2017pwj, Gair:2017ynp, Babak:2017tow}. Such detections would provide an important opportunity to probe the geometry of spacetime around supermassive black holes and test possible quantum gravity effects through precise GW observations.

\section{Discussions and Conclusions}\label{section5}

The gravitational waves from compact objects orbiting black holes furnish a powerful probe of strong-field gravity and the structure of spacetime. In particular, EMRIs are expected to be among the most informative sources for space-based detectors such as LISA, as they encode precise information about the background geometry through the orbital motion of the small body. Motivated by the above arguments, we have explored how modifications to spacetime arising from noncommutative geometry influence the dynamics of particles around black holes and the resulting gravitational radiation. Understanding these effects is crucial for determining whether future gravitational wave observations can reveal signatures of quantum gravity or deviations from general relativity in the strong-field regime.

In particular, we investigate periodic orbits and their corresponding waveforms within the context of non-commutative geometry. From the geodesic equations for black holes, we can analytically solve the equations. Then we use a special taxonomy \cite{Levin:2008mq} to distinguish different types of periodic orbits in non-commutative geometry. In this scheme, each periodic orbit is described by a set of parameters $(z, w, v)$. This study examined the effect of the non-commutative parameter $\Theta$ on the orbits of particles around a black hole. Unlike the classic Schwarzschild case, the presence of $\Theta$ significantly altered these orbits.

The radiation of gravitational waves from periodic orbits in non-commutative geometry is preliminarily considered. These results may provide a way to distinguish between black holes in noncommutative geometry and the Schwarzschild black hole. We analyze an EMRI system consisting of a test object with mass $ m = 10 M_\odot $ following periodic orbits around an SMBH, having mass $M = 10^7 M_\odot$. Using the numerical kludge scheme, we investigated the resulting gravitational waveforms by positioning the system at a luminosity distance of $D_L = 2$ Gpc from the detector, with an inclination angle of $\iota = \pi/4$ and a longitude of pericenter  $\zeta = \pi/4 $. This study demonstrates a clear correlation between the gravitational waveforms emitted by a small object orbiting an SMBH and the object's zoom-whirl orbital behavior. Higher zoom numbers correspond to more complex waveform substructures. Furthermore, the presence of a $\Theta$ significantly impacts these waveforms. To assess the detectability of gravitational waves from EMRIs with periodic orbits, we analyzed their time-domain waveforms using discrete Fourier transforms to extract the frequency spectra. The results indicate that the frequencies of these gravitational waves generally fall within the sensitivity range of space-based detectors. From the spectra, we determined the characteristic strains and observed that, for certain combinations of $(z,w,v)$, the strains exceed the sensitivity threshold of LISA. This suggests that space-based gravitational wave observatories could detect signals from EMRIs with periodic orbits, providing a promising avenue for exploring supermassive black holes with dark matter halos.
Thus, our study highlights that the properties of $\Theta$ play a critical role in shaping GW signals, offering promising potential for future observations to probe the influence of non-commutativity in strong gravitational fields.

Here, we make a few remarks on the limitations of the waveform calculations and the potential extensions of the current study. First, we employ the adiabatic approximation, which neglects the backreaction of gravitational radiation on periodic orbits, a valid approximation when considering only a few orbital periods, as in this study. Exploring the impact of gravitational radiation on the long-term evolution of periodic orbits represents an interesting avenue for future research. Second, we also ignore the contributions of multipoles of higher than the quadratic order. It is also crucial to develop more accurate waveform models that include higher-order multipole moments in the gravitational wave expansion. 

In future work, we plan to extend this analysis in several ways. One important step will be to include the radiation reaction effects to study how gravitational wave emission changes the orbits over time. It will also be beneficial to extend the quadrupole approximation and incorporate higher multipole moments to achieve more accurate waveforms. Another natural extension is to consider rotating noncommutative black holes, where spin may further affect the orbital motion and emitted radiation. Finally, with future space-based detectors like LISA~\cite{LISA:2017pwj} and Taiji~\cite{Ruan:2018tsw}, improved waveform templates from such studies could help test deviations from general relativity and explore possible signatures of quantum gravity in the strong-field region. Finally, once these accurate waveforms become available, we will be able to investigate how future gravitational wave detectors might constrain or test the effects of dark matter on periodic orbits. We hope to address these challenges in future studies.

\section*{Acknowledgements}

This work is supported by the National Natural Science Foundation of China under Grants No. 12275238 and 11675143, the National Key Research and Development Program under Grant No. 2020YFC2201503, and the Zhejiang Provincial Natural Science Foundation of China under Grants No. LR21A050001 and No. LY20A050002, and the Fundamental Research Funds for the Provincial Universities of Zhejiang in China under Grant No. RF-A2019015.

%The data analyses and results visualization in this work made use of \texttt{BILBY} \cite{Romero-Shaw:2020owr, Ashton:2018jfp}, \texttt{dynesty} \cite{Speagle:2019ivv}, \texttt{LALSuite} \cite{LALSuite}, \texttt{Numpy} \cite{Harris:2020xlr, vanderWalt:2011bqk}, \texttt{Scipy} \cite{Virtanen:2019joe}, and \texttt{matplotlib} \cite{Hunter:2007ouj}.

\appendix

%\bibliography{sample}

\bibliographystyle{apsrev4-1}
\bibliography{main}

@article{Wei:2025qlh,
    author = "Wei, Ze-Lin and Zhang, Jing and Xie, Yi and Yin, Pei-Lin",
    title = "{Probing a one-loop quantum-corrected Schwarzschild spacetime with precessing and periodic motion}",
    doi = "10.1140/epjc/s10052-025-14437-x",
    journal = "Eur. Phys. J. C",
    volume = "85",
    number = "6",
    pages = "698",
    year = "2025"
}

@article{Bian:2025ifp,
    author = "Bian, Ligong and others",
    title = "{Gravitational wave cosmology}",
    eprint = "2505.19747",
    archivePrefix = "arXiv",
    primaryClass = "gr-qc",
    doi = "10.1007/s11433-025-2740-8",
    journal = "Sci. China Phys. Mech. Astron.",
    volume = "69",
    number = "1",
    pages = "210401",
    year = "2026"
}

@article{Ni:2024acg,
    author = "Ni, Wei-Tou",
    title = "{Space gravitational wave detection: Progress and outlook}",
    eprint = "2409.00927",
    archivePrefix = "arXiv",
    primaryClass = "gr-qc",
    doi = "10.1360/SSPMA-2024-0186",
    journal = "Sci. Sin. Phys. Mech. Astro.",
    volume = "54",
    number = "7",
    pages = "270402",
    year = "2024"
}

@article{Zahra:2025tdo,
    author = "Zahra, Tehreem and Shabbir, Oreeda and Majeed, Bushra and Jamil, Mubasher and Rayimbaev, Javlon and Shermatov, Abubakir",
    title = "{Gravitational wave radiation from periodic orbits and quasi-periodic oscillations in Einstein non-linear Maxwell-Yukawa black hole}",
    eprint = "2510.22761",
    archivePrefix = "arXiv",
    primaryClass = "gr-qc",
    month = "10",
    year = "2025"
}

@article{Li:2025eln,
    author = "Li, Guo-He and Qiao, Chen-Kai and Tao, Jun",
    title = "{Periodic orbits and their gravitational waves in EMRIs: supermassive black hole affected by galactic dark matter halos}",
    eprint = "2510.24989",
    archivePrefix = "arXiv",
    primaryClass = "gr-qc",
    month = "10",
    year = "2025"
}

@article{Chen:2025aqh,
    author = "Chen, Jiawei and Yang, Jinsong",
    title = "{Periodic orbits and gravitational waveforms in quantum-corrected black hole spacetimes}",
    eprint = "2505.02660",
    archivePrefix = "arXiv",
    primaryClass = "gr-qc",
    doi = "10.1140/epjc/s10052-025-14457-7",
    journal = "Eur. Phys. J. C",
    volume = "85",
    number = "7",
    pages = "726",
    year = "2025"
}

@article{Deng:2025wzz,
    author = "Deng, Weike and Long, Sheng and Tan, Qin and Jing, Jiliang",
    title = "{Gravitational waveforms from periodic orbits around a charged black hole with scalar hair}",
    eprint = "2510.24468",
    archivePrefix = "arXiv",
    primaryClass = "gr-qc",
    month = "10",
    year = "2025"
}

@article{Choudhury:2025qsh,
    author = "Choudhury, Sayantan and Hossain, Md Khalid and Bauyrzhan, Gulnur and Yerzhanov, Koblandy",
    title = "{Gravitational Wave Signatures of Periodic Motion near Higher-Derivative Einstein-{\AE}ther Black Holes}",
    eprint = "2507.00904",
    archivePrefix = "arXiv",
    primaryClass = "gr-qc",
    month = "7",
    year = "2025"
}

@article{Li:2025sfe,
    author = "Li, Yong-Zhuang and Kuang, Xiao-Mei",
    title = "{The bound orbits and gravitational waveforms of timelike particles around renormalization group improved Kerr black holes}",
    eprint = "2509.07333",
    archivePrefix = "arXiv",
    primaryClass = "gr-qc",
    month = "9",
    year = "2025"
}

@article{Alloqulov:2025bxh,
    author = "Alloqulov, Mirzabek and Shaymatov, Sanjar and Ahmedov, Bobomurat and Zhu, Tao",
    title = "{Regular black hole's impact on the gravitational waveforms from periodic orbits}",
    eprint = "2508.05245",
    archivePrefix = "arXiv",
    primaryClass = "gr-qc",
    month = "8",
    year = "2025"
}

@article{Gong:2025mne,
    author = "Gong, Huajie and Long, Sheng and Wang, Xi-Jing and Xia, Zhongwu and Wu, Jian-Pin and Pan, Qiyuan",
    title = "{Gravitational waveforms from periodic orbits around a novel regular black hole}",
    eprint = "2509.23318",
    archivePrefix = "arXiv",
    primaryClass = "gr-qc",
    month = "9",
    year = "2025"
}

@article{Zare:2025aek,
    author = "Zare, Soroush and Zhu, Tao and Nieto, Luis M. and Lu, Shuo and Hassanabadi, Hassan",
    title = "{Probing regular black holes with sub-Planckian curvature through periodic orbits and their gravitational wave radiation}",
    eprint = "2510.05166",
    archivePrefix = "arXiv",
    primaryClass = "gr-qc",
    month = "10",
    year = "2025"
}

@article{Lu:2025cxx,
    author = "Lu, Shuo and Zhu, Tao",
    title = "{Gravitational radiations from periodic orbits around Einstein-{\AE}ther black holes}",
    eprint = "2505.00294",
    archivePrefix = "arXiv",
    primaryClass = "gr-qc",
    doi = "10.1016/j.dark.2025.102141",
    journal = "Phys. Dark Univ.",
    volume = "50",
    pages = "102141",
    year = "2025"
}

@article{LIGOScientific:2016aoc,
    author = "Abbott, B. P. and others",
    collaboration = "LIGO Scientific, Virgo",
    title = "{Observation of Gravitational Waves from a Binary Black Hole Merger}",
    eprint = "1602.03837",
    archivePrefix = "arXiv",
    primaryClass = "gr-qc",
    reportNumber = "LIGO-P150914",
    doi = "10.1103/PhysRevLett.116.061102",
    journal = "Phys. Rev. Lett.",
    volume = "116",
    number = "6",
    pages = "061102",
    year = "2016"
}

@article{LIGOScientific:2016vbw,
    author = "Abbott, B. P. and others",
    collaboration = "LIGO Scientific, Virgo",
    title = "{GW150914: First results from the search for binary black hole coalescence with Advanced LIGO}",
    eprint = "1602.03839",
    archivePrefix = "arXiv",
    primaryClass = "gr-qc",
    reportNumber = "LIGO-P1500269",
    doi = "10.1103/PhysRevD.93.122003",
    journal = "Phys. Rev. D",
    volume = "93",
    number = "12",
    pages = "122003",
    year = "2016"
}

@article{LIGOScientific:2016vlm,
    author = "Abbott, B. P. and others",
    collaboration = "LIGO Scientific, Virgo",
    title = "{Properties of the Binary Black Hole Merger GW150914}",
    eprint = "1602.03840",
    archivePrefix = "arXiv",
    primaryClass = "gr-qc",
    reportNumber = "LIGO-P1500218",
    doi = "10.1103/PhysRevLett.116.241102",
    journal = "Phys. Rev. Lett.",
    volume = "116",
    number = "24",
    pages = "241102",
    year = "2016"
}

@article{LIGOScientific:2016emj,
    author = "Abbott, B. P. and others",
    collaboration = "LIGO Scientific, Virgo",
    title = "{GW150914: The Advanced LIGO Detectors in the Era of First Discoveries}",
    eprint = "1602.03838",
    archivePrefix = "arXiv",
    primaryClass = "gr-qc",
    reportNumber = "LIGO-P1500237",
    doi = "10.1103/PhysRevLett.116.131103",
    journal = "Phys. Rev. Lett.",
    volume = "116",
    number = "13",
    pages = "131103",
    year = "2016"
}

@article{Levin:2008mq,
    author = "Levin, Janna and Perez-Giz, Gabe",
    title = "{A Periodic Table for Black Hole Orbits}",
    eprint = "0802.0459",
    archivePrefix = "arXiv",
    primaryClass = "gr-qc",
    doi = "10.1103/PhysRevD.77.103005",
    journal = "Phys. Rev. D",
    volume = "77",
    pages = "103005",
    year = "2008"
}

@article{Levin:2009sk,
    author = "Levin, Janna",
    title = "{Energy Level Diagrams for Black Hole Orbits}",
    eprint = "0907.5195",
    archivePrefix = "arXiv",
    primaryClass = "gr-qc",
    doi = "10.1088/0264-9381/26/23/235010",
    journal = "Class. Quant. Grav.",
    volume = "26",
    pages = "235010",
    year = "2009"
}

@article{Misra:2010pu,
    author = "Misra, Vedant and Levin, Janna",
    title = "{Rational Orbits around Charged Black Holes}",
    eprint = "1007.2699",
    archivePrefix = "arXiv",
    primaryClass = "gr-qc",
    doi = "10.1103/PhysRevD.82.083001",
    journal = "Phys. Rev. D",
    volume = "82",
    pages = "083001",
    year = "2010"
}

@article{Babar:2017gsg,
    author = "Babar, Gulmina Zaman and Babar, Adil Zaman and Lim, Yen-Kheng",
    title = "{Periodic orbits around a spherically symmetric naked singularity}",
    eprint = "1710.09581",
    archivePrefix = "arXiv",
    primaryClass = "gr-qc",
    doi = "10.1103/PhysRevD.96.084052",
    journal = "Phys. Rev. D",
    volume = "96",
    number = "8",
    pages = "084052",
    year = "2017"
}

@article{Hu:2017mde,
    author = "Hu, Wen-Rui and Wu, Yue-Liang",
    title = "{The Taiji Program in Space for gravitational wave physics and the nature of gravity}",
    doi = "10.1093/nsr/nwx116",
    journal = "Natl. Sci. Rev.",
    volume = "4",
    number = "5",
    pages = "685--686",
    year = "2017"
}

@article{TianQin:2015yph,
    author = "Luo, Jun and others",
    collaboration = "TianQin",
    title = "{TianQin: a space-borne gravitational wave detector}",
    eprint = "1512.02076",
    archivePrefix = "arXiv",
    primaryClass = "astro-ph.IM",
    doi = "10.1088/0264-9381/33/3/035010",
    journal = "Class. Quant. Grav.",
    volume = "33",
    number = "3",
    pages = "035010",
    year = "2016"
}

@article{Gong:2021gvw,
    author = "Gong, Yungui and Luo, Jun and Wang, Bin",
    title = "{Concepts and status of Chinese space gravitational wave detection projects}",
    eprint = "2109.07442",
    archivePrefix = "arXiv",
    primaryClass = "astro-ph.IM",
    doi = "10.1038/s41550-021-01480-3",
    journal = "Nature Astron.",
    volume = "5",
    number = "9",
    pages = "881--889",
    year = "2021"
}

@article{Danzmann:1997hm,
    author = "Danzmann, K.",
    title = "{LISA: An ESA cornerstone mission for a gravitational wave observatory}",
    doi = "10.1088/0264-9381/14/6/002",
    journal = "Class. Quant. Grav.",
    volume = "14",
    pages = "1399--1404",
    year = "1997"
}

@article{Schutz:1999xj,
    author = "Schutz, Bernard F.",
    title = "{Gravitational wave astronomy}",
    eprint = "gr-qc/9911034",
    archivePrefix = "arXiv",
    reportNumber = "AEI-1999-34",
    doi = "10.1088/0264-9381/16/12A/307",
    journal = "Class. Quant. Grav.",
    volume = "16",
    pages = "A131--A156",
    year = "1999"
}

@article{Gair:2004iv,
    author = "Gair, Jonathan R. and Barack, Leor and Creighton, Teviet and Cutler, Curt and Larson, Shane L. and Phinney, E. Sterl and Vallisneri, Michele",
    title = "{Event rate estimates for LISA extreme mass ratio capture sources}",
    eprint = "gr-qc/0405137",
    archivePrefix = "arXiv",
    doi = "10.1088/0264-9381/21/20/003",
    journal = "Class. Quant. Grav.",
    volume = "21",
    pages = "S1595--S1606",
    year = "2004"
}

@article{LISA:2017pwj,
    author = "Amaro-Seoane, Pau and others",
    collaboration = "LISA",
    title = "{Laser Interferometer Space Antenna}",
    eprint = "1702.00786",
    archivePrefix = "arXiv",
    primaryClass = "astro-ph.IM",
    month = "2",
    year = "2017"
}

@article{Maselli:2021men,
    author = "Maselli, Andrea and Franchini, Nicola and Gualtieri, Leonardo and Sotiriou, Thomas P. and Barsanti, Susanna and Pani, Paolo",
    title = "{Detecting fundamental fields with LISA observations of gravitational waves from extreme mass-ratio inspirals}",
    eprint = "2106.11325",
    archivePrefix = "arXiv",
    primaryClass = "gr-qc",
    doi = "10.1038/s41550-021-01589-5",
    journal = "Nature Astron.",
    volume = "6",
    number = "4",
    pages = "464--470",
    year = "2022"
}

@article{Levin:2008ci,
    author = "Levin, Janna and Grossman, Becky",
    title = "{Dynamics of Black Hole Pairs. I. Periodic Tables}",
    eprint = "0809.3838",
    archivePrefix = "arXiv",
    primaryClass = "gr-qc",
    doi = "10.1103/PhysRevD.79.043016",
    journal = "Phys. Rev. D",
    volume = "79",
    pages = "043016",
    year = "2009"
}

@article{Bambhaniya:2020zno,
    author = "Bambhaniya, Parth and Solanki, Divyesh N. and Dey, Dipanjan and Joshi, Ashok B. and Joshi, Pankaj S. and Patel, Vishva",
    title = "{Precession of timelike bound orbits in Kerr spacetime}",
    eprint = "2007.12086",
    archivePrefix = "arXiv",
    primaryClass = "gr-qc",
    doi = "10.1140/epjc/s10052-021-08997-x",
    journal = "Eur. Phys. J. C",
    volume = "81",
    number = "3",
    pages = "205",
    year = "2021"
}

@article{Rana:2019bsn,
    author = "Rana, Prerna and Mangalam, A.",
    title = "{Astrophysically relevant bound trajectories around a Kerr black hole}",
    eprint = "1901.02730",
    archivePrefix = "arXiv",
    primaryClass = "gr-qc",
    doi = "10.1088/1361-6382/ab004c",
    journal = "Class. Quant. Grav.",
    volume = "36",
    pages = "045009",
    year = "2019"
}

@article{Liu:2018vea,
    author = "Liu, Changqing and Ding, Chikun and Jing, Jiliang",
    title = "{Periodic orbits around Kerr Sen black holes}",
    eprint = "1804.05883",
    archivePrefix = "arXiv",
    primaryClass = "gr-qc",
    doi = "10.1088/0253-6102/71/12/1461",
    journal = "Commun. Theor. Phys.",
    volume = "71",
    number = "12",
    pages = "1461",
    year = "2019"
}

@article{Lin:2023rmo,
    author = "Lin, Hou-Yu and Deng, Xue-Mei",
    title = "{Precessing and periodic orbits around hairy black holes in Horndeski{\textquoteright}s Theory}",
    doi = "10.1140/epjc/s10052-023-11487-x",
    journal = "Eur. Phys. J. C",
    volume = "83",
    number = "4",
    pages = "311",
    year = "2023"
}

@article{Yao:2023ziq,
    author = "Yao, Jin-Tao and Li, Xin",
    title = "{Closed orbits in axial symmetric Finslerian extension of a Schwarzschild black hole}",
    doi = "10.1103/PhysRevD.108.084067",
    journal = "Phys. Rev. D",
    volume = "108",
    number = "8",
    pages = "084067",
    year = "2023"
}

@article{Lin:2022llz,
    author = "Lin, Hou-Yu and Deng, Xue-Mei",
    title = "{Bound Orbits and Epicyclic Motions around Renormalization Group Improved Schwarzschild Black Holes}",
    doi = "10.3390/universe8050278",
    journal = "Universe",
    volume = "8",
    number = "5",
    pages = "278",
    year = "2022"
}

@article{Chan:2025ocy,
    author = "Chan, Zoe C. S. and Lim, Yen-Kheng",
    title = {{Periodic orbits of neutral test particles in Reissner{\textendash}Nordstr{\"o}m naked singularities}},
    eprint = "2502.03082",
    archivePrefix = "arXiv",
    primaryClass = "gr-qc",
    doi = "10.1007/s10714-025-03368-3",
    journal = "Gen. Rel. Grav.",
    volume = "57",
    number = "2",
    pages = "35",
    year = "2025"
}

@article{Wang:2022tfo,
    author = "Wang, Ruifang and Gao, Fabao and Chen, Huixiang",
    title = "{Periodic orbits around a static spherically symmetric black hole surrounded by quintessence}",
    doi = "10.1016/j.aop.2022.169167",
    journal = "Annals Phys.",
    volume = "447",
    number = "1",
    pages = "169167",
    year = "2022"
}

@article{Lin:2023eyd,
    author = "Lin, Hou-Yu and Deng, Xue-Mei",
    title = "{Dynamics of test particles around hairy black holes in Horndeski{\textquoteright}s theory}",
    doi = "10.1016/j.aop.2023.169360",
    journal = "Annals Phys.",
    volume = "455",
    pages = "169360",
    year = "2023"
}

@article{Habibina:2022ztd,
    author = "Habibina, A. S. and Ramadhan, H. S.",
    title = "{Bound orbits around charged black strings}",
    eprint = "2205.14635",
    archivePrefix = "arXiv",
    primaryClass = "gr-qc",
    doi = "10.1016/j.aop.2022.169169",
    journal = "Annals Phys.",
    volume = "448",
    pages = "169169",
    year = "2023"
}

@article{Zhang:2022psr,
    author = "Zhang, Jing and Xie, Yi",
    title = "{Probing a self-complete and Generalized-Uncertainty-Principle black hole with precessing and periodic motion}",
    doi = "10.1007/s10509-022-04046-5",
    journal = "Astrophys. Space Sci.",
    volume = "367",
    number = "2",
    pages = "17",
    year = "2022"
}

@article{Lin:2022wda,
    author = "Lin, Hou-Yu and Deng, Xue-Mei",
    title = "{Precessing and periodic orbits around Lee{\textendash}Wick black holes}",
    doi = "10.1140/epjp/s13360-022-02391-6",
    journal = "Eur. Phys. J. Plus",
    volume = "137",
    number = "2",
    pages = "176",
    year = "2022"
}

@article{Gao:2021arw,
    author = "Gao, Bo and Deng, Xue-Mei",
    title = "{Bound orbits around modified Hayward black holes}",
    doi = "10.1142/S0217732321502370",
    journal = "Mod. Phys. Lett. A",
    volume = "36",
    number = "33",
    pages = "2150237",
    year = "2021"
}

@article{Lin:2021noq,
    author = "Lin, Hou-Yu and Deng, Xue-Mei",
    title = "{Rational orbits around 4 $\mathcal D$ Einstein{\textendash}Lovelock black holes}",
    doi = "10.1016/j.dark.2020.100745",
    journal = "Phys. Dark Univ.",
    volume = "31",
    pages = "100745",
    year = "2021"
}

@article{Deng:2020yfm,
    author = "Deng, Xue-Mei",
    title = "{Geodesics and periodic orbits around quantum-corrected black holes}",
    doi = "10.1016/j.dark.2020.100629",
    journal = "Phys. Dark Univ.",
    volume = "30",
    pages = "100629",
    year = "2020"
}

@article{Zhou:2020zys,
    author = "Zhou, Tian-Yi and Xie, Yi",
    title = "{Precessing and periodic motions around a black-bounce/traversable wormhole}",
    doi = "10.1140/epjc/s10052-020-08661-w",
    journal = "Eur. Phys. J. C",
    volume = "80",
    number = "11",
    pages = "1070",
    year = "2020"
}

@article{Wang:2025wob,
    author = "Wang, Chao-Hui and Zhang, Yu-Peng and Zhu, Tao and Wei, Shao-Wen",
    title = "{A new type of multi-branch periodic orbits in dyonic black holes}",
    eprint = "2508.20558",
    archivePrefix = "arXiv",
    primaryClass = "gr-qc",
    month = "8",
    year = "2025"
}

@article{Gao:2020wjz,
    author = "Gao, Bo and Deng, Xue-Mei",
    title = "{Bound orbits around Bardeen black holes}",
    doi = "10.1016/j.aop.2020.168194",
    journal = "Annals Phys.",
    volume = "418",
    pages = "168194",
    year = "2020"
}

@article{Deng:2020hxw,
    author = "Deng, Xue-Mei",
    title = "{Periodic orbits around brane-world black holes}",
    doi = "10.1140/epjc/s10052-020-8067-7",
    journal = "Eur. Phys. J. C",
    volume = "80",
    number = "6",
    pages = "489",
    year = "2020"
}

@article{Wei:2019zdf,
    author = "Wei, Shao-Wen and Yang, Jie and Liu, Yu-Xiao",
    title = "{Geodesics and periodic orbits in Kehagias-Sfetsos black holes in deformed Hor̆ava-Lifshitz gravity}",
    eprint = "1904.03129",
    archivePrefix = "arXiv",
    primaryClass = "gr-qc",
    doi = "10.1103/PhysRevD.99.104016",
    journal = "Phys. Rev. D",
    volume = "99",
    number = "10",
    pages = "104016",
    year = "2019"
}

@article{Pugliese:2013xfa,
    author = "Pugliese, Daniela and Quevedo, Hernando and Ruffini, Remo",
    title = {{General classification of charged test particle circular orbits in Reissner--Nordstr{\"o}m spacetime}},
    eprint = "1304.2940",
    archivePrefix = "arXiv",
    primaryClass = "gr-qc",
    doi = "10.1140/epjc/s10052-017-4769-x",
    journal = "Eur. Phys. J. C",
    volume = "77",
    number = "4",
    pages = "206",
    year = "2017"
}

@article{Healy:2009zm,
    author = "Healy, James and Levin, Janna and Shoemaker, Deirdre",
    title = "{Zoom-Whirl Orbits in Black Hole Binaries}",
    eprint = "0907.0671",
    archivePrefix = "arXiv",
    primaryClass = "gr-qc",
    doi = "10.1103/PhysRevLett.103.131101",
    journal = "Phys. Rev. Lett.",
    volume = "103",
    pages = "131101",
    year = "2009"
}

@article{Zhang:2022zox,
    author = "Zhang, Jing and Xie, Yi",
    title = {{Probing a black-bounce-Reissner{\textendash}Nordstr{\"o}m spacetime with precessing and periodic motion}},
    doi = "10.1140/epjc/s10052-022-10846-4",
    journal = "Eur. Phys. J. C",
    volume = "82",
    number = "10",
    pages = "854",
    year = "2022"
}

@article{Haroon:2025rzx,
    author = "Haroon, Sumarna and Zhu, Tao",
    title = "{Periodic orbits and their gravitational wave radiations in a black hole with a dark matter halo}",
    eprint = "2502.09171",
    archivePrefix = "arXiv",
    primaryClass = "gr-qc",
    doi = "10.1103/ckdt-wtsl",
    journal = "Phys. Rev. D",
    volume = "112",
    number = "4",
    pages = "044046",
    year = "2025"
}

@article{Tu:2023xab,
    author = "Tu, Ze-Yi and Zhu, Tao and Wang, Anzhong",
    title = "{Periodic orbits and their gravitational wave radiations in a polymer black hole in loop quantum gravity}",
    eprint = "2304.14160",
    archivePrefix = "arXiv",
    primaryClass = "gr-qc",
    doi = "10.1103/PhysRevD.108.024035",
    journal = "Phys. Rev. D",
    volume = "108",
    number = "2",
    pages = "024035",
    year = "2023"
}

@article{Azreg-Ainou:2020bfl,
    author = {Azreg-A{\"\i}nou, Mustapha and Chen, Zihang and Deng, Bojun and Jamil, Mubasher and Zhu, Tao and Wu, Qiang and Lim, Yen-Kheng},
    title = "{Orbital mechanics and quasiperiodic oscillation resonances of black holes in Einstein-{\AE}ther theory}",
    eprint = "2004.02602",
    archivePrefix = "arXiv",
    primaryClass = "gr-qc",
    doi = "10.1103/PhysRevD.102.044028",
    journal = "Phys. Rev. D",
    volume = "102",
    number = "4",
    pages = "044028",
    year = "2020"
}

@article{Yang:2024lmj,
    author = "Yang, Sen and Zhang, Yu-Peng and Zhu, Tao and Zhao, Li and Liu, Yu-Xiao",
    title = "{Gravitational waveforms from periodic orbits around a quantum-corrected black hole}",
    eprint = "2407.00283",
    archivePrefix = "arXiv",
    primaryClass = "gr-qc",
    doi = "10.1088/1475-7516/2025/01/091",
    journal = "JCAP",
    volume = "01",
    pages = "091",
    year = "2025"
}

@article{Shabbir:2025kqh,
    author = {Shabbir, Oreeda and Jamil, Mubasher and Azreg-A{\"\i}nou, Mustapha},
    title = "{Periodic orbits and their gravitational wave radiations around the Schwarzschild-MOG black hole}",
    eprint = "2501.04367",
    archivePrefix = "arXiv",
    primaryClass = "gr-qc",
    doi = "10.1016/j.dark.2025.101816",
    journal = "Phys. Dark Univ.",
    volume = "47",
    pages = "101816",
    year = "2025"
}

@article{Junior:2024tmi,
    author = "Junior, Ednaldo L. B. and Junior, Jos{\'e} Tarciso S. S. and Lobo, Francisco S. N. and Rodrigues, Manuel E. and Rubiera-Garcia, Diego and da Silva, Lu{\'\i}s F. Dias and Vieira, Henrique A.",
    title = "{Periodical orbits and waveforms with spontaneous Lorentz symmetry-breaking in Kalb{\textendash}Ramond gravity}",
    eprint = "2412.00769",
    archivePrefix = "arXiv",
    primaryClass = "gr-qc",
    doi = "10.1140/epjc/s10052-025-14299-3",
    journal = "Eur. Phys. J. C",
    volume = "85",
    number = "5",
    pages = "557",
    year = "2025"
}

@article{Zhao:2024exh,
    author = "Zhao, Lai and Tang, Meirong and Xu, Zhaoyi",
    title = "{Periodic orbits and gravitational wave radiation in short hair black hole spacetimes for an extreme mass ratio system}",
    eprint = "2411.01979",
    archivePrefix = "arXiv",
    primaryClass = "gr-qc",
    doi = "10.1140/epjc/s10052-025-13767-0",
    journal = "Eur. Phys. J. C",
    volume = "85",
    number = "1",
    pages = "36",
    year = "2025"
}

@article{Jiang:2024cpe,
    author = "Jiang, Hanyu and Alloqulov, Mirzabek and Wu, Qiang and Shaymatov, Sanjar and Zhu, Tao",
    title = "{Periodic orbits and plasma effects on gravitational weak lensing by self-dual black hole in loop quantum gravity}",
    doi = "10.1016/j.dark.2024.101627",
    journal = "Phys. Dark Univ.",
    volume = "46",
    pages = "101627",
    year = "2024"
}

@article{Yang:2024cnd,
    author = "Yang, Sen and Zhang, Yu-Peng and Zhu, Tao and Zhao, Li and Liu, Yu-Xiao",
    title = "{Constraining polymerized black holes with quasi-circular extreme mass-ratio inspirals*}",
    eprint = "2412.04302",
    archivePrefix = "arXiv",
    primaryClass = "gr-qc",
    doi = "10.1088/1674-1137/adef1a",
    journal = "Chin. Phys.",
    volume = "49",
    number = "11",
    pages = "115107",
    year = "2025"
}

@article{Meng:2024cnq,
    author = "Meng, Liping and Xu, Zhaoyi and Tang, Meirong",
    title = "{Bound orbits and gravitational wave radiation around the hairy black hole}",
    eprint = "2411.01858",
    archivePrefix = "arXiv",
    primaryClass = "gr-qc",
    doi = "10.1140/epjc/s10052-025-14032-0",
    journal = "Eur. Phys. J. C",
    volume = "85",
    number = "3",
    pages = "306",
    year = "2025"
}

@article{Li:2024tld,
    author = "Li, Yong-Zhuang and Kuang, Xiao-Mei and Sang, Yu",
    title = "{Precessing and periodic timelike orbits and their potential applications in Einsteinian cubic gravity}",
    eprint = "2401.16071",
    archivePrefix = "arXiv",
    primaryClass = "gr-qc",
    doi = "10.1140/epjc/s10052-024-12895-3",
    journal = "Eur. Phys. J. C",
    volume = "84",
    number = "5",
    pages = "529",
    year = "2024"
}

@article{QiQi:2024dwc,
    author = "Qi, Qi and Kuang, Xiao-Mei and Li, Yong-Zhuang and Sang, Yu",
    title = "{Timelike bound orbits and pericenter precession around black hole with conformally coupled scalar hair}",
    eprint = "2407.01958",
    archivePrefix = "arXiv",
    primaryClass = "gr-qc",
    doi = "10.1140/epjc/s10052-024-12989-y",
    journal = "Eur. Phys. J. C",
    volume = "84",
    number = "6",
    pages = "645",
    year = "2024"
}

@article{Alloqulov:2025ucf,
    author = "Alloqulov, Mirzabek and Xamidov, Tursunali and Shaymatov, Sanjar and Ahmedov, Bobomurat",
    title = "{Gravitational waveforms from periodic orbits around a Schwarzschild black hole embedded in a Dehnen-type dark matter halo}",
    eprint = "2504.05236",
    archivePrefix = "arXiv",
    primaryClass = "gr-qc",
    doi = "10.1140/epjc/s10052-025-14529-8",
    journal = "Eur. Phys. J. C",
    volume = "85",
    number = "7",
    pages = "798",
    year = "2025"
}

@article{Wang:2025hla,
    author = "Wang, Chao-Hui and Meng, Xiang-Cheng and Zhang, Yu-Peng and Zhu, Tao and Wei, Shao-Wen",
    title = "{Equatorial periodic orbits and gravitational waveforms in a black hole free of Cauchy horizon}",
    eprint = "2502.08994",
    archivePrefix = "arXiv",
    primaryClass = "gr-qc",
    doi = "10.1088/1475-7516/2025/07/021",
    month = "2",
    year = "2025"
}

@article{Babak:2006uv,
    author = "Babak, Stanislav and Fang, Hua and Gair, Jonathan R. and Glampedakis, Kostas and Hughes, Scott A.",
    title = "{'Kludge' gravitational waveforms for a test-body orbiting a Kerr black hole}",
    eprint = "gr-qc/0607007",
    archivePrefix = "arXiv",
    doi = "10.1103/PhysRevD.75.024005",
    journal = "Phys. Rev. D",
    volume = "75",
    pages = "024005",
    year = "2007",
    note = "[Erratum: Phys.Rev.D 77, 04990 (2008)]"
}

@article{Thorne:1980ru,
    author = "Thorne, K. S.",
    title = "{Multipole Expansions of Gravitational Radiation}",
    doi = "10.1103/RevModPhys.52.299",
    journal = "Rev. Mod. Phys.",
    volume = "52",
    pages = "299--339",
    year = "1980"
}

@article{Hamil:2024ppj,
    author = {Hamil, B. and L{\"u}tf{\"u}o{\u{g}}lu, B. C.},
    title = "{Noncommutative Schwarzschild black hole surrounded by quintessence: Thermodynamics, Shadows and Quasinormal modes}",
    eprint = "2401.09295",
    archivePrefix = "arXiv",
    primaryClass = "gr-qc",
    doi = "10.1016/j.dark.2024.101484",
    journal = "Phys. Dark Univ.",
    volume = "44",
    pages = "101484",
    year = "2024"
}

@article{Nicolini:2005vd,
    author = "Nicolini, Piero and Smailagic, Anais and Spallucci, Euro",
    title = "{Noncommutative geometry inspired Schwarzschild black hole}",
    eprint = "gr-qc/0510112",
    archivePrefix = "arXiv",
    doi = "10.1016/j.physletb.2005.11.004",
    journal = "Phys. Lett. B",
    volume = "632",
    pages = "547--551",
    year = "2006"
}

@article{Nicolini:2005gy,
    author = "Nicolini, Piero",
    editor = "Mankoc Borstnik, Norma and Nielsen, Holger Bech and Froggatt, Colin D. and Lukman, Dragan",
    title = "{Noncommutative nonsingular black holes}",
    eprint = "hep-th/0510203",
    archivePrefix = "arXiv",
    journal = "Bled Workshops Phys.",
    volume = "6",
    number = "2",
    pages = "79--87",
    year = "2005"
}

@article{Kumar:2017hgs,
    author = "Kumar, Rahul and Ghosh, Sushant G.",
    title = "{Accretion onto a noncommutative geometry inspired black hole}",
    eprint = "1703.10479",
    archivePrefix = "arXiv",
    primaryClass = "gr-qc",
    reportNumber = "EPJC-S10052-017-5141-X",
    doi = "10.1140/epjc/s10052-017-5141-x",
    journal = "Eur. Phys. J. C",
    volume = "77",
    number = "9",
    pages = "577",
    year = "2017"
}

@article{Ghosh:2017odw,
    author = "Ghosh, Sushant G.",
    title = "{Noncommutative geometry inspired Einstein{\textendash}Gauss{\textendash}Bonnet black holes}",
    eprint = "1707.08174",
    archivePrefix = "arXiv",
    primaryClass = "gr-qc",
    doi = "10.1088/1361-6382/aaaead",
    journal = "Class. Quant. Grav.",
    volume = "35",
    number = "8",
    pages = "085008",
    year = "2018"
}

@article{Ghosh:2020cob,
    author = "Ghosh, Sushant G. and Maharaj, Sunil D.",
    title = "{Noncommutative inspired black holes in regularized 4D Einstein{\textendash}Gauss{\textendash}Bonnet theory}",
    eprint = "2004.13519",
    archivePrefix = "arXiv",
    primaryClass = "gr-qc",
    doi = "10.1016/j.dark.2021.100793",
    journal = "Phys. Dark Univ.",
    volume = "31",
    pages = "100793",
    year = "2021"
}

@article{Hughes:1999bq,
    author = "Hughes, Scott A.",
    title = "{The Evolution of circular, nonequatorial orbits of Kerr black holes due to gravitational wave emission}",
    eprint = "gr-qc/9910091",
    archivePrefix = "arXiv",
    doi = "10.1103/PhysRevD.65.069902",
    journal = "Phys. Rev. D",
    volume = "61",
    number = "8",
    pages = "084004",
    year = "2000",
    note = "[Erratum: Phys.Rev.D 63, 049902 (2001), Erratum: Phys.Rev.D 65, 069902 (2002), Erratum: Phys.Rev.D 67, 089901 (2003), Erratum: Phys.Rev.D 78, 109902 (2008), Erratum: Phys.Rev.D 90, 109904 (2014)]"
}

@article{Isoyama:2021jjd,
    author = "Isoyama, Soichiro and Fujita, Ryuichi and Chua, Alvin J. K. and Nakano, Hiroyuki and Pound, Adam and Sago, Norichika",
    title = "{Adiabatic Waveforms from Extreme-Mass-Ratio Inspirals: An Analytical Approach}",
    eprint = "2111.05288",
    archivePrefix = "arXiv",
    primaryClass = "gr-qc",
    reportNumber = "YITP-21-114,KUNS-2901,OCU-PHYS-551,AP-GR-175, YITP-21-114",
    doi = "10.1103/PhysRevLett.128.231101",
    journal = "Phys. Rev. Lett.",
    volume = "128",
    number = "23",
    pages = "231101",
    year = "2022"
}

@article{Kiselev:2002dx,
    author = "Kiselev, V. V.",
    title = "{Quintessence and black holes}",
    eprint = "gr-qc/0210040",
    archivePrefix = "arXiv",
    primaryClass = "gr-qc",
    doi = "10.1088/0264-9381/20/6/310",
    journal = "Class. Quant. Grav.",
    volume = "20",
    pages = "1187--1198",
    year = "2003"
}

@article{Giri:2006rc,
    author = "Giri, Prabir R.",
    title = "{Noncommutative geometry inspired black holes and regular distributions}",
    eprint = "hep-th/0604188",
    archivePrefix = "arXiv",
    primaryClass = "hep-th",
    doi = "10.1142/S0217751X07036245",
    journal = "Int. J. Mod. Phys. A",
    volume = "22",
    pages = "2047--2056",
    year = "2007"
}

@article{Anacleto:2019tdj,
    author = "Anacleto, M. A. and Brito, F. A. and Campos, J. A. V. and Passos, E.",
    title = "{Noncommutative inspired black holes with smeared mass distributions}",
    eprint = "1907.13107",
    archivePrefix = "arXiv",
    primaryClass = "hep-th",
    doi = "10.1016/j.physletb.2020.135334",
    journal = "Phys. Lett. B",
    volume = "803",
    pages = "135334",
    year = "2020"
}

@article{Levin:2008yp,
    author = "Levin, Janna and Perez-Giz, Gabe",
    title = "{Homoclinic Orbits around Spinning Black Holes. I. Exact Solution for the Kerr Separatrix}",
    eprint = "0811.3814",
    archivePrefix = "arXiv",
    primaryClass = "gr-qc",
    doi = "10.1103/PhysRevD.79.124013",
    journal = "Phys. Rev. D",
    volume = "79",
    pages = "124013",
    year = "2009"
}

@book{Chandrasekhar:1985kt,
    author = "Chandrasekhar, Subrahmanyan",
    title = "{The mathematical theory of black holes}",
    isbn = "978-0-19-850370-5",
    year = "1985"
}

@article{Barack:2003fp,
    author = "Barack, Leor and Cutler, Curt",
    title = "{LISA capture sources: Approximate waveforms, signal-to-noise ratios, and parameter estimation accuracy}",
    eprint = "gr-qc/0310125",
    archivePrefix = "arXiv",
    doi = "10.1103/PhysRevD.69.082005",
    journal = "Phys. Rev. D",
    volume = "69",
    pages = "082005",
    year = "2004"
}

@article{Hughes:2000ssa,
    author = "Hughes, Scott A.",
    editor = "Schutz, Bernard F.",
    title = "{Gravitational waves from extreme mass ratio inspirals: Challenges in mapping the space-time of massive, compact objects}",
    eprint = "gr-qc/0008058",
    archivePrefix = "arXiv",
    doi = "10.1088/0264-9381/18/19/314",
    journal = "Class. Quant. Grav.",
    volume = "18",
    pages = "4067--4074",
    year = "2001"
}

@article{Gair:2017ynp,
    author = "Gair, Jonathan R. and Babak, Stanislav and Sesana, Alberto and Amaro-Seoane, Pau and Barausse, Enrico and Berry, Christopher P. L. and Berti, Emanuele and Sopuerta, Carlos",
    editor = "Giardini, Domencio and Jetzer, Philippe",
    title = "{Prospects for observing extreme-mass-ratio inspirals with LISA}",
    eprint = "1704.00009",
    archivePrefix = "arXiv",
    primaryClass = "astro-ph.GA",
    doi = "10.1088/1742-6596/840/1/012021",
    journal = "J. Phys. Conf. Ser.",
    volume = "840",
    number = "1",
    pages = "012021",
    year = "2017"
}

@article{Babak:2017tow,
    author = "Babak, Stanislav and Gair, Jonathan and Sesana, Alberto and Barausse, Enrico and Sopuerta, Carlos F. and Berry, Christopher P. L. and Berti, Emanuele and Amaro-Seoane, Pau and Petiteau, Antoine and Klein, Antoine",
    title = "{Science with the space-based interferometer LISA. V: Extreme mass-ratio inspirals}",
    eprint = "1703.09722",
    archivePrefix = "arXiv",
    primaryClass = "gr-qc",
    doi = "10.1103/PhysRevD.95.103012",
    journal = "Phys. Rev. D",
    volume = "95",
    number = "10",
    pages = "103012",
    year = "2017"
}

@article{Ruan:2018tsw,
    author = "Ruan, Wen-Hong and Guo, Zong-Kuan and Cai, Rong-Gen and Zhang, Yuan-Zhong",
    title = "{Taiji program: Gravitational-wave sources}",
    eprint = "1807.09495",
    archivePrefix = "arXiv",
    primaryClass = "gr-qc",
    doi = "10.1142/S0217751X2050075X",
    journal = "Int. J. Mod. Phys. A",
    volume = "35",
    number = "17",
    pages = "2050075",
    year = "2020"
}

@article{AraujoFilho:2025rwr,
    author = "Ara{\'u}jo Filho, A. A.",
    title = "{Particle production induced by a Lorentzian non-commutative spacetime}",
    eprint = "2502.19366",
    archivePrefix = "arXiv",
    primaryClass = "gr-qc",
    doi = "10.1016/j.aop.2025.170167",
    journal = "Annals Phys.",
    volume = "481",
    pages = "170167",
    year = "2025"
}

@article{AraujoFilho:2025jcu,
    author = "Ara{\'u}jo Filho, A. A. and Heidari, N. and Lobo, Iarley P.",
    title = "{A non-commutative Kalb-Ramond black hole}",
    eprint = "2507.17390",
    archivePrefix = "arXiv",
    primaryClass = "gr-qc",
    doi = "10.1088/1475-7516/2025/09/076",
    journal = "JCAP",
    volume = "09",
    pages = "076",
    year = "2025"
}

@article{Filho:2024zxx,
    author = {Filho, A. A. Ara{\'u}jo and Nascimento, J. R. and Petrov, A. Yu. and Porf{\'\i}rio, P. J. and {\"O}vg{\"u}n, Ali},
    title = "{Effects of non-commutative geometry on black hole properties}",
    eprint = "2406.12015",
    archivePrefix = "arXiv",
    primaryClass = "gr-qc",
    doi = "10.1016/j.dark.2024.101630",
    journal = "Phys. Dark Univ.",
    volume = "46",
    pages = "101630",
    year = "2024"
}

@mastersthesis{Depies:2009im,
    author = "Depies, Matthew R",
    title = "{Gravitational Waves and Light Cosmic Strings}",
    eprint = "0908.3680",
    archivePrefix = "arXiv",
    primaryClass = "gr-qc",
    type = "Other thesis",
    month = "8",
    year = "2009"
}

@article{Larson:1999we,
    author = "Larson, Shane L. and Hiscock, William A. and Hellings, Ronald W.",
    title = "{Sensitivity curves for spaceborne gravitational wave interferometers}",
    eprint = "gr-qc/9909080",
    archivePrefix = "arXiv",
    reportNumber = "MSUPHY-99-04",
    doi = "10.1103/PhysRevD.62.062001",
    journal = "Phys. Rev. D",
    volume = "62",
    pages = "062001",
    year = "2000"
}

\end{document}